\documentclass[journal,10pt]{IEEEtran}
\usepackage{cite}
\usepackage{graphicx}
\usepackage{amsmath}
\usepackage{array}
\usepackage{mdwmath}
\usepackage{amssymb}
\usepackage{mdwtab}
\usepackage{stfloats}
\usepackage[tight,footnotesize]{subfigure}
\usepackage{amsmath,amsthm}
\usepackage{threeparttable}
\usepackage{color}
\usepackage{epstopdf}

\newtheorem{theorem}{\it \textbf{Theorem}}

\newtheorem{lemma}{\it  \textbf{Lemma}}

\newtheorem{assumption}{\it  \textbf{Assumption}}
\begin{document}
	\title{ Impact of Rotary-Wing UAV Wobbling on Millimeter-wave Air-to-Ground Wireless Channel}
	\author{Songjiang Yang,~\textit{Graduate~Student~Member,~IEEE}, Zitian Zhang, \\ Jiliang Zhang,~\textit{Senior~Member,~IEEE},  Jie Zhang,~\textit{Senior~Member,~IEEE}
	\thanks{Manuscript received XXX, XX, 2015; revised XXX, XX, 2015.}
	\thanks{Songjiang Yang and Jiliang Zhang are with the Department of Electronic and Electrical Engineering, the University of Sheffield, Sheffield, S10 2TN, UK, (e-mail: syang16@sheffield.ac.uk, jiliang.zhang@sheffield.ac.uk).
		
	Zitian Zhang is with Ranplan Wireless Network Design Ltd., Cambridge, CB23 3UY, UK.		
	
	Jie Zhang is with the Department of Electronic and Electrical Engineering, the University of Sheffield, Sheffield, S10 2TN, UK, and also with Ranplan Wireless Network Design Ltd., Cambridge, CB23 3UY, UK.}}
	\markboth{IEEE TRANSACTIONS ON VEHICULAR TECHNOLOGY,~Vol.~X, No.~X, Month~20XX}
	{Shell \MakeLowercase{\textit{et al.}}: Bare Demo of IEEEtran.cls for Journals}
	\maketitle
	\begin{abstract}		
	\noindent
	Millimeter-wave rotary-wing (RW) unmanned aerial vehicle (UAV) air-to-ground (A2G) links face unpredictable Doppler effect arising from the inevitable wobbling of RW UAV.
	Moreover, the time-varying channel characteristics during transmission lead to inaccurate channel estimation, which in turn results in the deteriorated bit error probability performance of the UAV A2G link.
	This paper studies the impact of mechanical wobbling on the Doppler effect of the millimeter-wave wireless channel between a hovering RW UAV and a ground node. 
	Our contributions of this paper lie in: i) modeling the wobbling process of a hovering RW UAV; ii) developing an analytical model to derive the channel temporal autocorrelation function (ACF) for the millimeter-wave RW UAV A2G link in a closed-form expression; and iii) investigating how RW UAV wobbling impacts the Doppler effect on the millimeter-wave RW UAV A2G link.
	Numerical results show that different RW UAV wobbling patterns impact the amplitude and the frequency of ACF oscillation in the millimeter-wave RW UAV A2G link. 
	For UAV wobbling, the channel temporal ACF decreases quickly and the impact of the Doppler effect is significant on the millimeter-wave A2G link.
	
	\end{abstract}
	\IEEEpeerreviewmaketitle
	\section{Introduction}
	In the 5G era, unmanned aerial vehicle (UAV) air-to-ground (A2G) links have been widely investigated to facilitate numerous applications such as flexible coverage and capacity enhancements, emergency assistance, disaster relief, etc. \cite{Liu2019IOT, Wang2020TVT}. 
	Moreover, the integration of UAV and millimeter-wave communications has been proposed to provide a high data rate UAV A2G link to support these applications\cite{Dabiri2020TWC}.
	The miniaturization of the aerial base station for UAV carrying could be realized due to the small wavelength in millimeter-wave frequency \cite{Rappaport2013ACCESS}.  
	The integration of a high beam directive antenna is a novel solution to provide the high-capacity communication links \cite{Zhang2020TVT, Zhu2020JSAC, Xiao2020CC}.
	
	However, in an UAV A2G link, UAV hovering in the air may experience mechanical wobbling in millimeter scale owing to imperfect mechanical control and various environmental issues, such as wind gusts, bad weather, and high vibration frequency of their propellers and rotors \cite{Banagar2020TVT, Plasencia2012IJARS}.
	In other words, UAV can be seen as mobile transceivers with non-negligible velocity in hovering status, which leads to a severe Doppler effect in the millimeter-wave RW UAV A2G link \cite{Xiao2016COM, Zhang2019WCZ}. 
	Moreover, the time-varying channel characteristics in A2G link lead to outdated channel state information from the channel estimation, resulting in deteriorated bit error probability (BEP) performance of the UAV A2G link. \cite{Li2009CL}.
	Therefore, to reap the benefits of millimeter-wave RW UAV communications, the wireless channel for millimeter-wave UAV A2G link should be distinctly characterized with regard to the Doppler effect.

	\subsection{Related Works}
	To this end, the research on millimeter-wave RW UAV A2G propagation channel is still in its infancy.
	The reliable analytical A2G channel model approaches, which can be categorized as deterministic models and geometry-based stochastic models (GBSMs), are necessary to characterize the propagation behavior of the millimeter-wave UAV channel \cite{Khawaja2019TUT,Cheng2019IOT}.
	The deterministic method can depict the realistic behavior of the electromagnetic wave propagation by ray-tracing software with high accuracy \cite{Khuwaja2018TUT}. 
	The GBSM can obtain the spatial-temporal channel characteristics in a geometric simulated environment, with low computational complexity \cite{Cheng2019IOT}.
	For instance, in \cite{Jiang2018CL, Zhu2019IETM, Jia2019IETC}, the 3D geometry-based UAV-multi-input-multi-output (MIMO) A2G channel model was proposed to indicate the space-time correlation function of the high mobility fixed-wing (FW) UAV, where GBSMs could be used to derive the channel characteristics, such as the temporal autocorrelation function (ACF), the Doppler power spectrum density (PSD), and spatial correlation.   
	The mmWave FW UAV-MIMO GBSM, which used the birth-death process to model the non-stationary	property of scatterers in mmWave UAV A2G link, was proposed in \cite{Michailidis2019TVT}. 
	Since the FW UAV requires continuous stable high-speed mobility, the Doppler effect in the FW UAV A2G link can be analyzed and compensated by using the space-time correlation function for non-stationary UAV scenarios \cite{Sharma2019TWC}.
	However, the Doppler effect of the RW UAV brought by mechanical wobbling at hovering status cannot be analyzed by mentioned channel model method due to the intrinsic randomness of the RW UAV wobbling \cite{Plasencia2012IJARS}.
		
	A handful of UAV A2G research works have considered the wobbling of RW UAV \cite{Dabiri2020TWC, Banagar2020TVT, Hou2019VTC}.
	In \cite{Dabiri2020TWC}, the authors discussed the antenna mismatch of the transmitter and the receiver caused by wobbling in the millimeter-wave RW UAV A2G link.
	In \cite{Banagar2020TVT}, the authors studied the impact of UAV wobbling on the coherence time of the wireless channel using Rician fading, where the assumption of wobbling process is too ideal for the practical environment. 
	In \cite{Hou2019VTC}, the authors analyzed the impact of propellers' rotation on the Doppler shift of RW UAV A2G channel.
	
	To summarize, none of the state-of-the-art RW UAV A2G channel models has studied the Doppler effect brought by the wavelength-scale mechanical wobbling at hovering status in millimeter-wave bands. 
	\subsection{Contribution}
	In this paper, we present the first attempt to analytically investigate the impact of UAV mechanical wobbling on the Doppler effect in millimeter-wave RW UAV A2G links. 
	Specifically, the channel temporal ACF and the Doppler PSD of millimeter-wave RW UAV channel under mechanical wobbling are derived. 
	The major contributions of this paper are summarized as follows:
	\begin{itemize}
	\item The RW UAV movement model at hovering status under mechanical wobbling is introduced.
	Two  parameters, i.e., the vibration frequency and the velocity envelope covariance, are defined to capture the characteristic of UAV mechanical wobbling.  
	\item The channel temporal ACF of millimeter-wave RW UAV A2G link is derived in a closed-form expression based on RW UAV mechanical wobbling movement model at hovering status. 
	The analytical expression of the channel temporal ACF is verified by Monte Carlo simulations. 
	\item The Doppler PSD of millimeter-wave RW UAV A2G link with mechanical wobbling is computed based on the analytical channel temporal ACF. 
	\item A key observation is that even for small UAV wobbling, the BEP performance of the UAV A2G link deteriorates quickly making the link difficult to establish a reliable communication link.
	\end{itemize} 
	
	The proposed analytical model of the channel temporal ACF and the Doppler PSD will be applied to predict the impact of the mechanical wobbling on the Doppler effect in millimeter-wave RW UAV A2G links. 
	
    The rest of this paper is organized as follows. 
	In Section II, the system model and assumptions of the RW UAV mechanical wobbling at hovering status in millimeter-wave UAV A2G link are presented. 
	In Section III, analytical results of the channel temporal ACF and the Doppler PSD in millimeter-wave RW UAV A2G link under mechanical wobbling are derived.
	In Section IV, numerical results are provided to verify our analytical expressions and impacts of UAV mechanical wobbling on millimeter-wave RW UAV channel at hovering status are discussed.
	Finally, in Section V, our main conclusions and some future research directions are drawn.
	\section{System Model}
	\begin{figure}[t]
		\centering
		\includegraphics[width=0.95\linewidth]{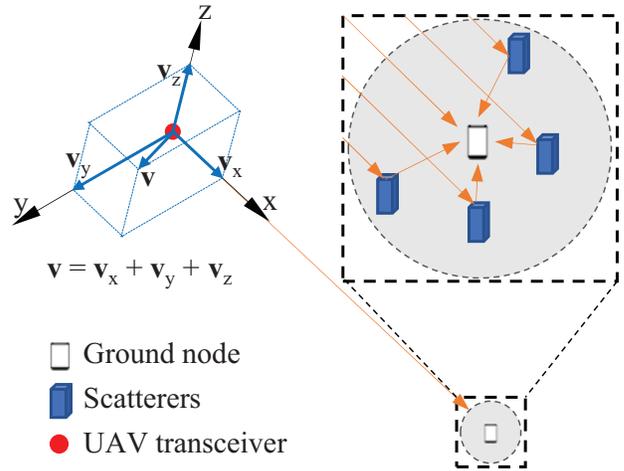}
		\caption{The millimeter-wave UAV A2G link with UAV mechanical wobbling.}
		\label{fig:uavsystemmodel}
	\end{figure}

\begin{figure*} [t]
	\centering
	\subfigure[$\mu=40,\ \omega_{\mathrm{v}}=20\pi$]{\includegraphics[width=0.43\linewidth]{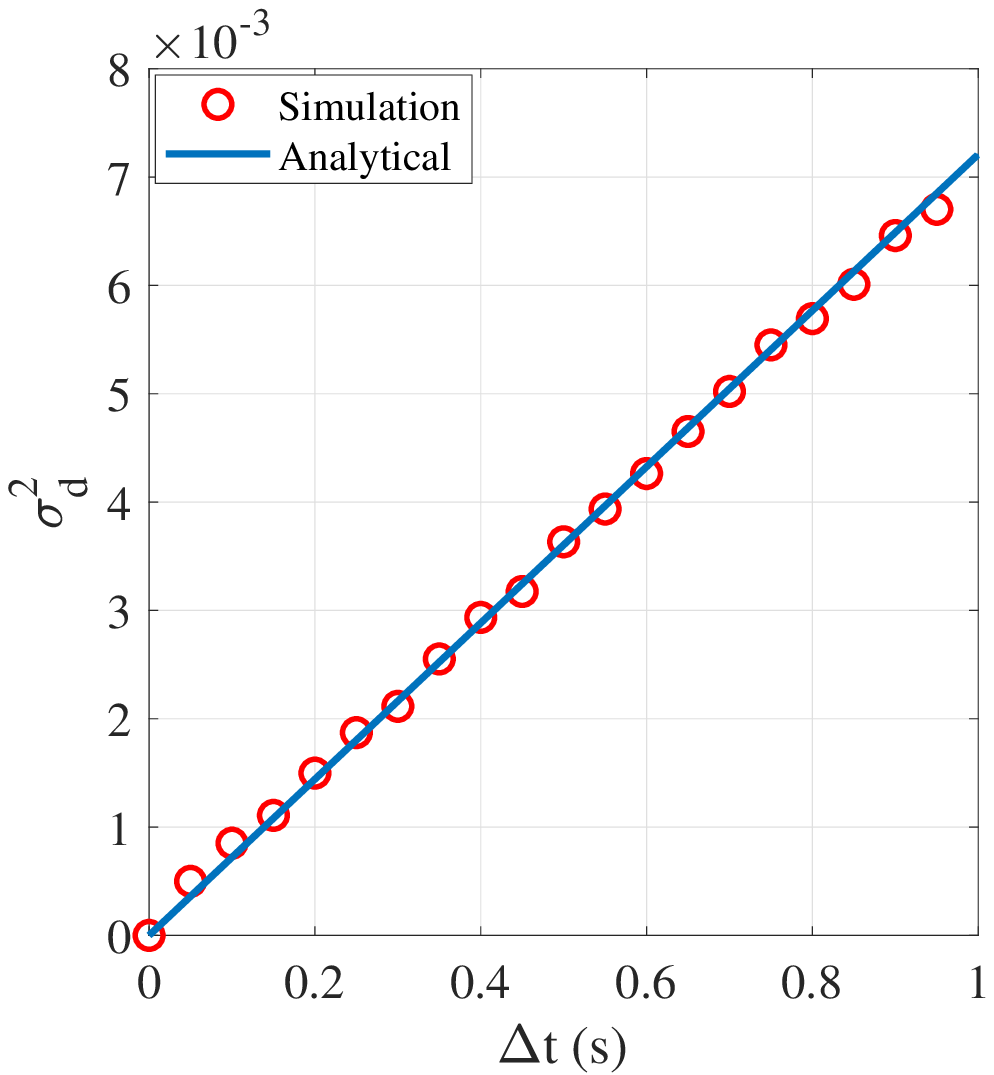}}
	\subfigure[$\mu=0,\ \omega_{\mathrm{v}}=0$]{\includegraphics[width=0.43\linewidth]{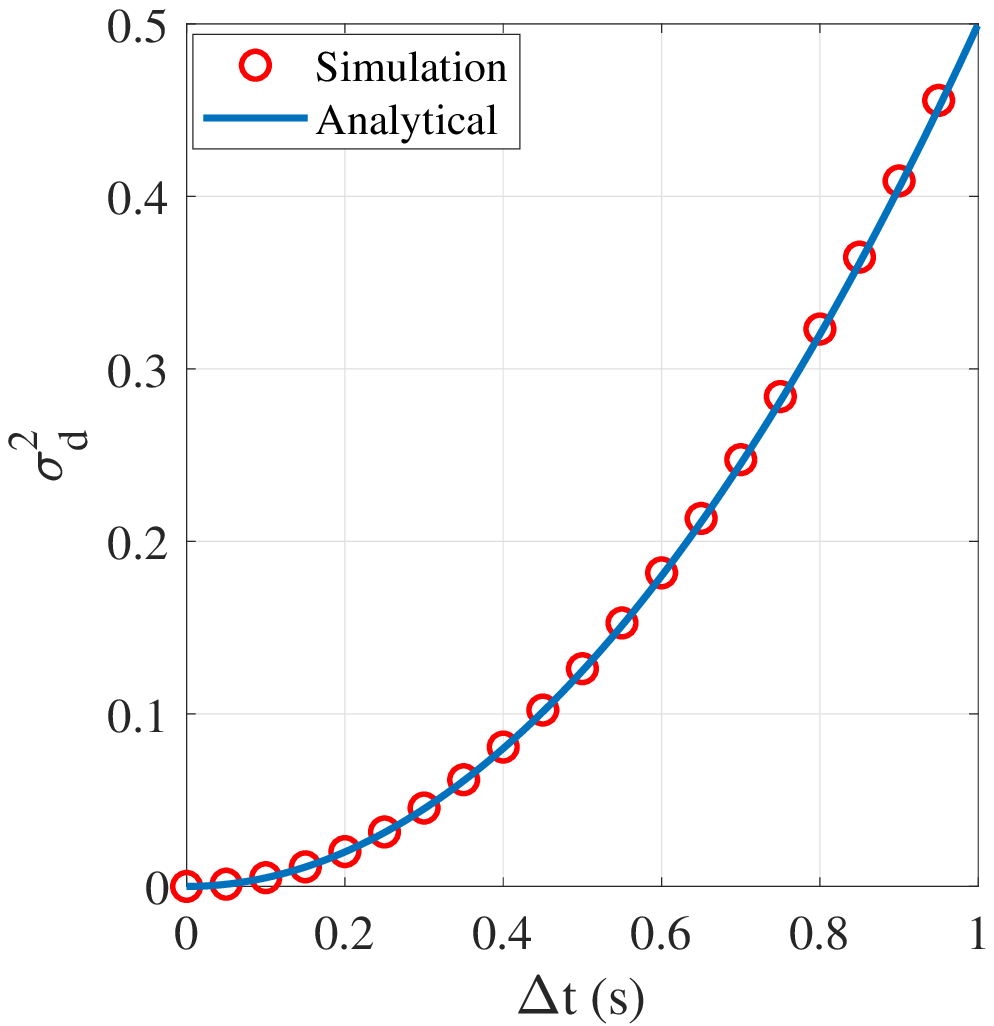}}\\
	\centering
	\subfigure[$\mu=40,\ \omega_{\mathrm{v}}=0$]{\includegraphics[width=0.43\linewidth]{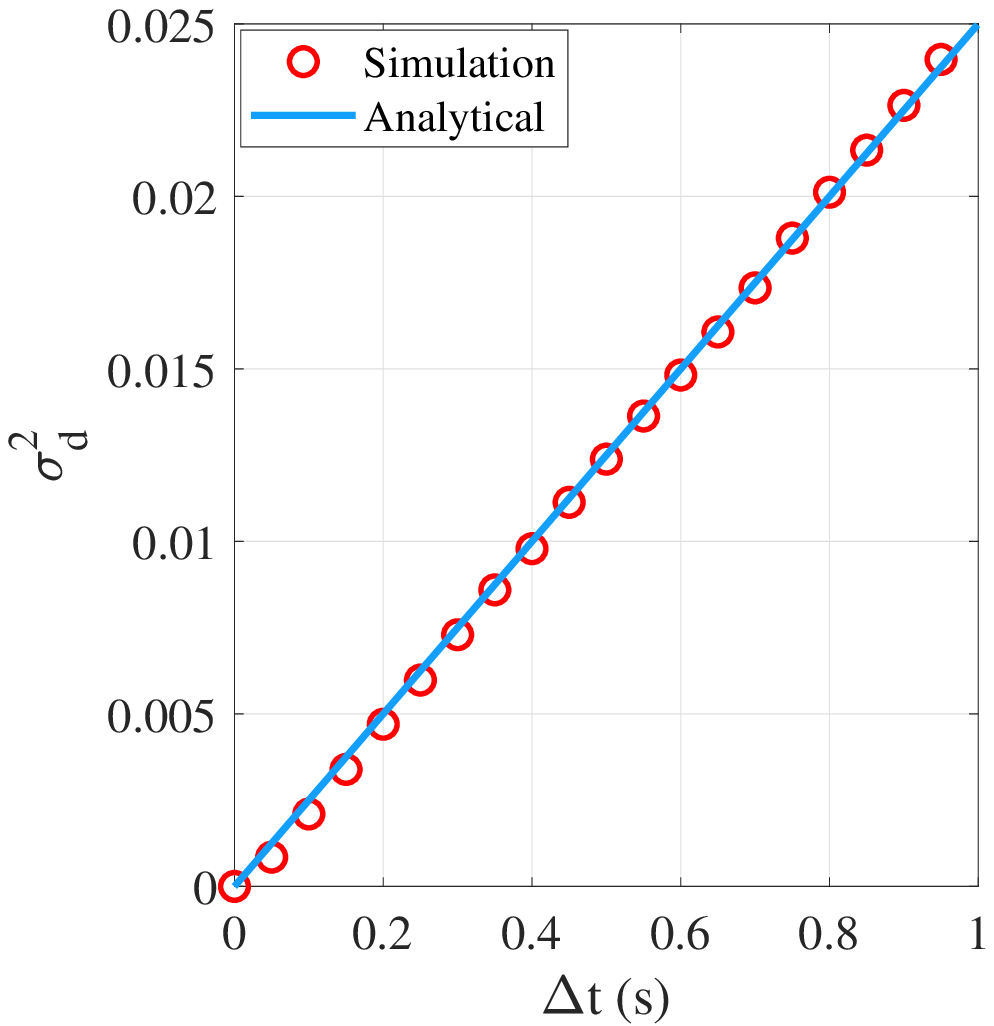}}
	\subfigure[$\mu=0,\ \omega_{\mathrm{v}}=20\pi$]{\includegraphics[width=0.43\linewidth]{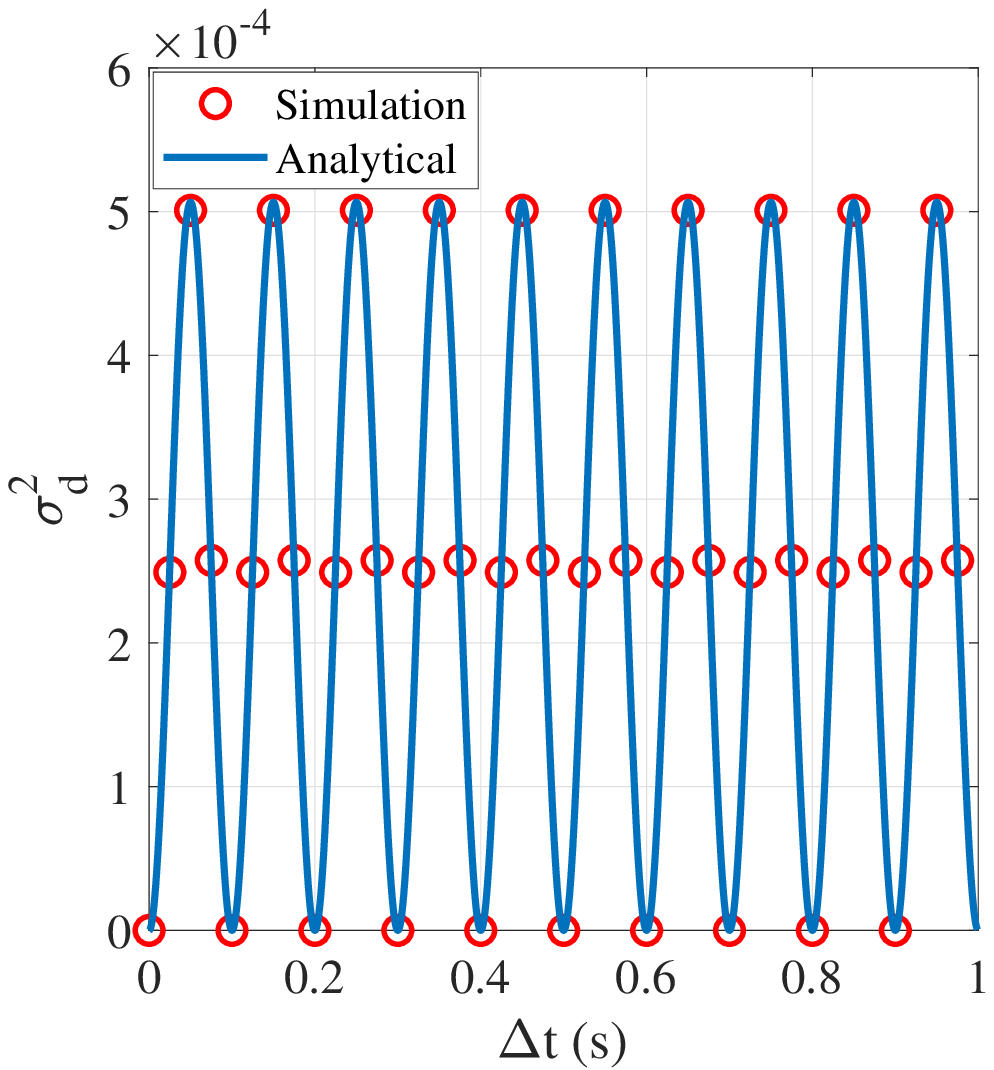}}
	\caption{The analytical and simulation $\sigma_{\mathrm{d}}^2$ at different cases with $\sigma_{\mathrm{v}}=1$. The time separation of movement caused by  motor vibration will be  in millisecond scale.  Solid lines illustrate analytical results and markers show simulation results.}
	\label{fig:sigma_d}
\end{figure*}

	\begin{figure*}[t]
	\centering
	\subfigure[$\mu =0,\ \omega_{\mathrm{v}}=0$]{\includegraphics[width=0.43\linewidth]{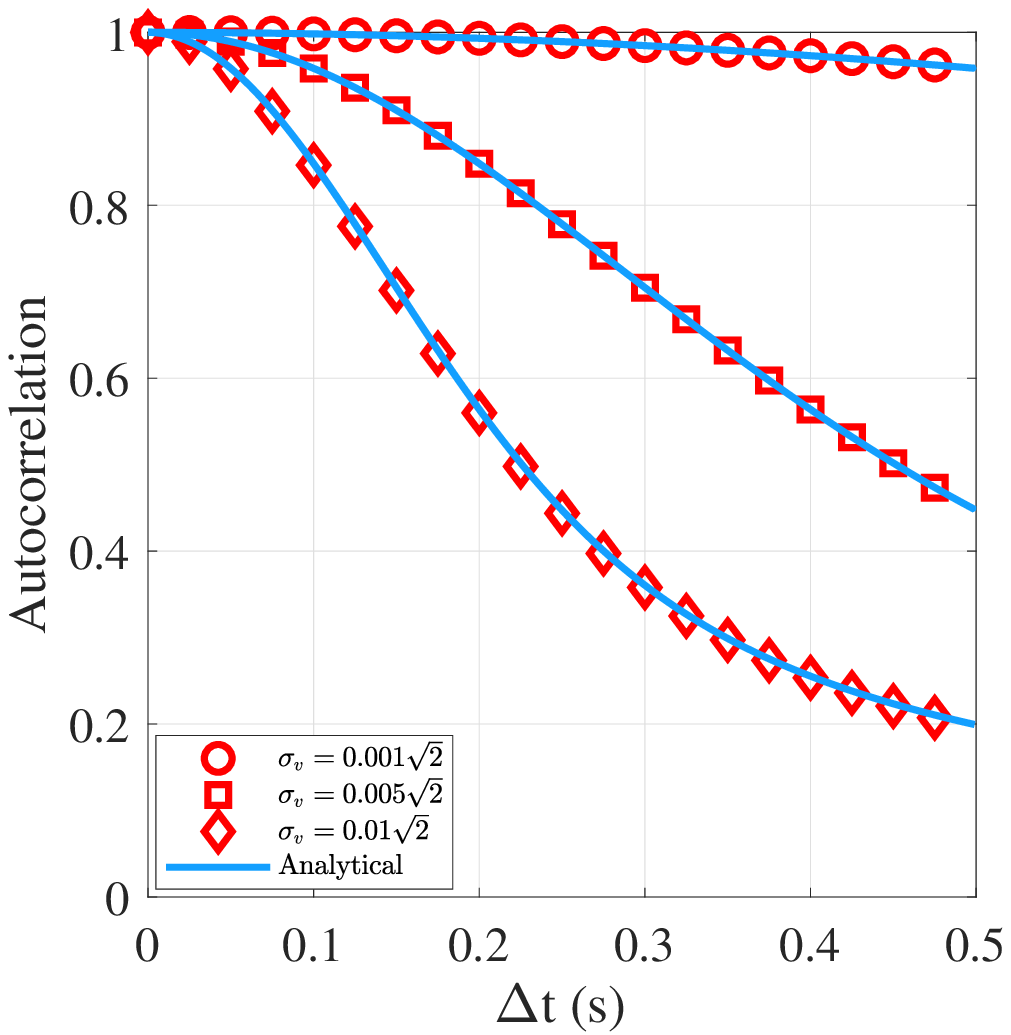}}
	\subfigure[$\mu =0,\ \omega_{\mathrm{v}}=20\pi$]{\includegraphics[width=0.43\linewidth]{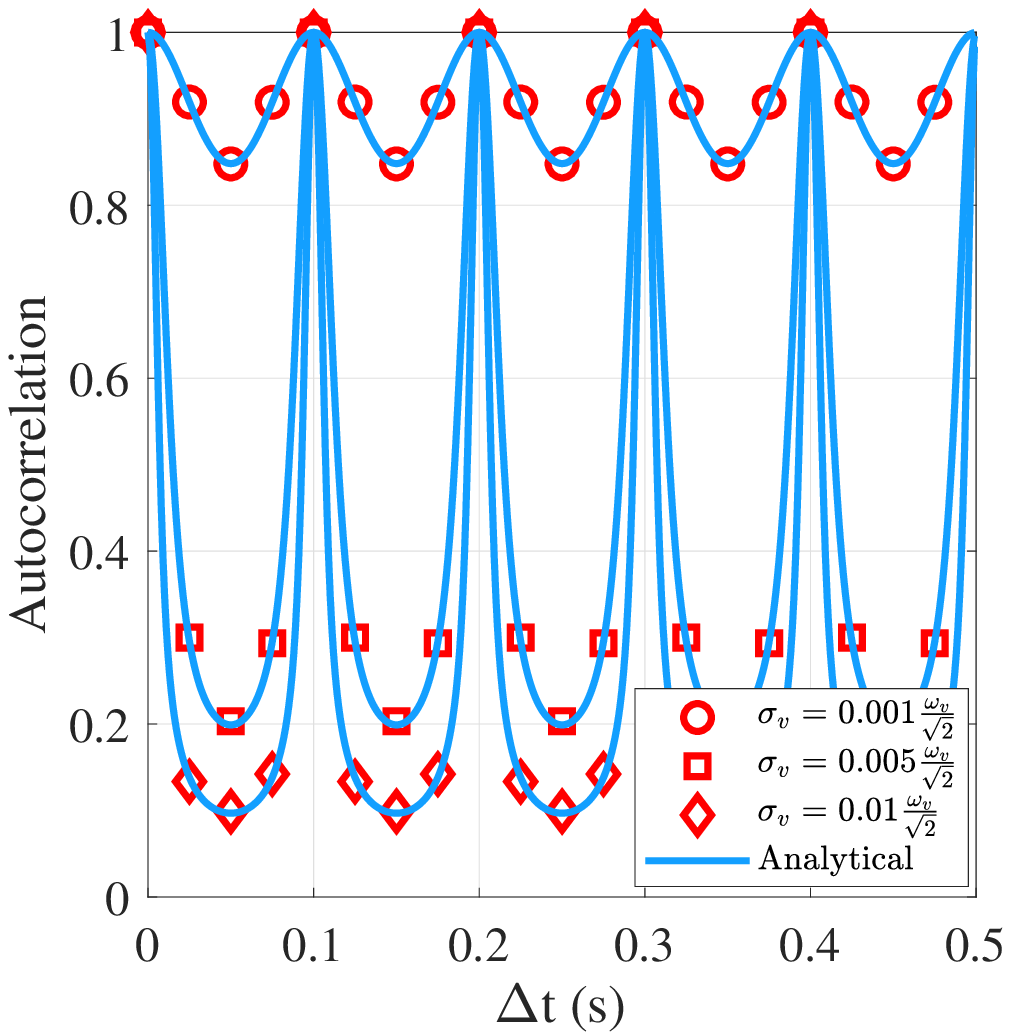}}
	\subfigure[$\mu =30,\ \omega_{\mathrm{v}}=0$]{\includegraphics[width=0.43\linewidth]{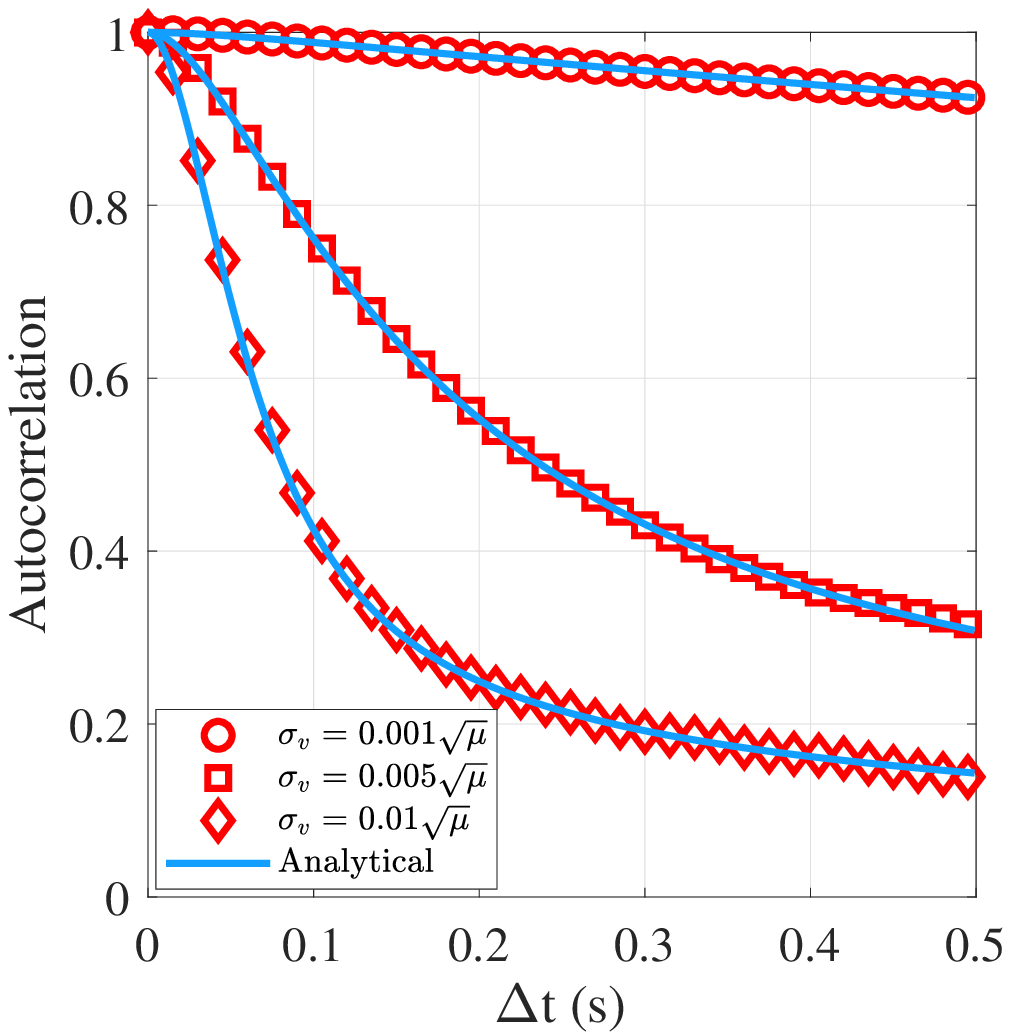}}
	\subfigure[$\mu =30,\ \omega_{\mathrm{v}}=20\pi$]{\includegraphics[width=0.43\linewidth]{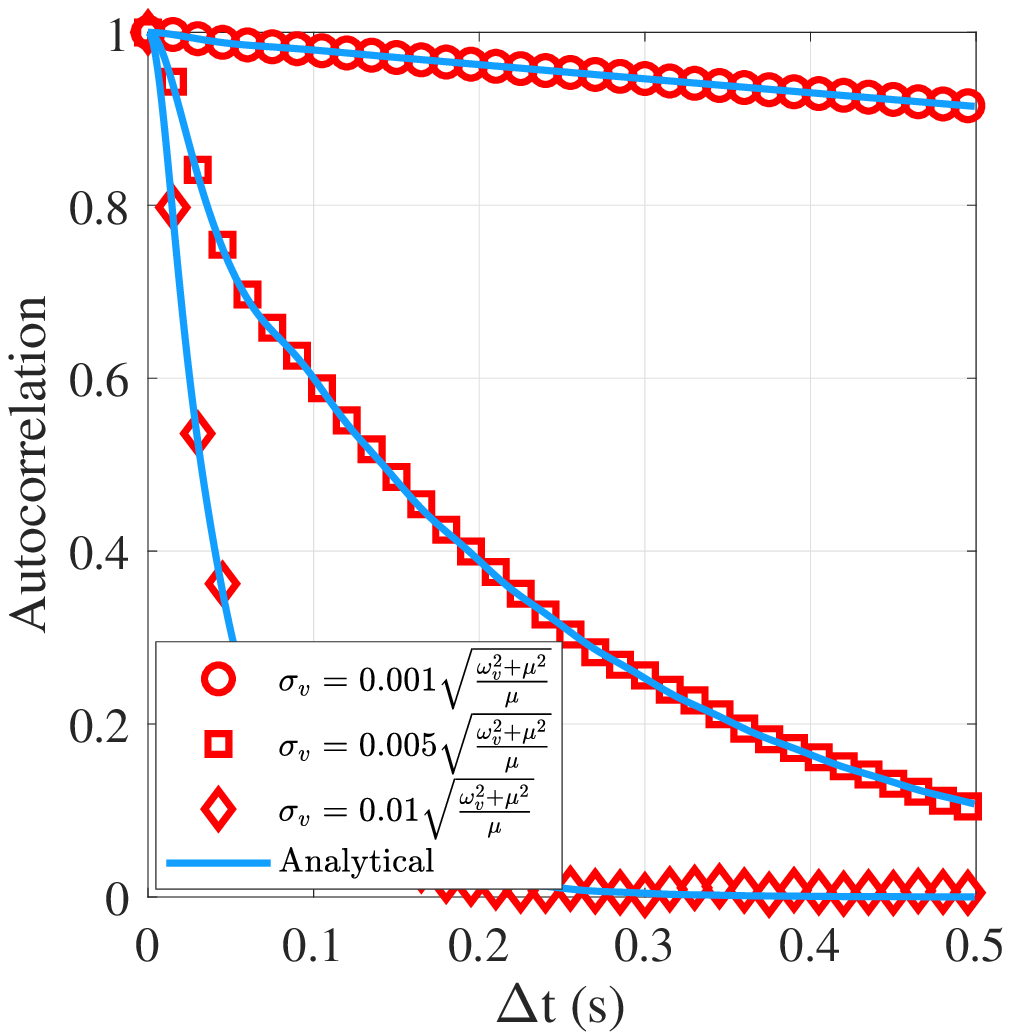}}
	\caption{The impact of $\sigma_{\mathrm{v}}$ on channel temporal ACF. Solid lines illustrate analytical results and markers show simulation results.}
	\label{fig:ACF_sigmad}		
\end{figure*}
	In this section, we first introduce the system model and assumptions in millimeter-wave RW UAV A2G link.
	Then, we derive the UAV wobbling movement process of the millimeter-wave RW UAV A2G link. 
	
	In this study, we design a millimeter-wave UAV A2G link between a hovering UAV and a ground node.
	As shown in Fig. \ref{fig:uavsystemmodel}, the UAV is equipped with a single horn directional antenna (transceiver) located at the bottom of the UAV platform.
	The directional horn antenna supports the directional beam to overcome high attenuation of the millimeter-wave propagation \cite{He2020VTM, Zhang2019ACCESS}.  
	According to Fig. \ref{fig:uavsystemmodel}, the velocity of the UAV mechanical wobbling is denoted as $ \mathbf{v} $, which could be superposed by following a Cartesian coordinate system, i.e., $\mathbf{v}=\mathbf{v}_\mathrm{x}+\mathbf{v}_\mathrm{y}+\mathbf{v}_\mathrm{z}$. 
	Moreover, one axis direction of Cartesian coordinate system is set as the same direction as the UAV A2G link to support the following analysis.
	More specifically, in Fig. \ref{fig:uavsystemmodel}, the x-axis is set as the same direction as the link and the radial velocity between the UAV and the ground node is $ \mathbf{v}_\mathrm{x} $.
	
	All key assumptions applied in this paper are listed as follows.
	\begin{assumption}
	The hovering height of RW UAV is much higher than those of building and other scatterers to keep the LoS wireless A2G link, as in \cite{Zhu2019CL}. 
	Moreover, the probability of the LoS link between RW UAV and BS is high and the dominant propagation mechanism is free space transmission \cite{Zeng2019PIEEE}.
	When the UAV is hovering at high altitude, for multipath channel model, scatterers nearby the ground node will be considered together. 
	Therefore, all links in this paper based on Assumption 1 are assumed as LoS links.
	\end{assumption}
	\begin{assumption}
	The velocity of the UAV mechanical wobbling is the superposition of the radial velocity and the tangential velocity in the millimeter-wave RW UAV A2G link, which is shown in Fig. \ref{fig:uavsystemmodel}. 
	The tangential velocity of the UAV wobbling may lead to variation of angle of arrival and departure in UAV A2G link, which could affect the Doppler effect in the UAV A2G channel model. 
	However, this effect could be negligible when the ground node position is fixed and the RW UAV is hovering at a high altitude \cite{Khawaja2018GSMM}, which are defined in Assumption 1.
	Therefore, the Doppler effect in our model is only caused by radial velocity of the RW UAV wobbling.
	We model the radial velocity, denoted by $V(t)$, to describe the wobbling movement of the RW UAV \cite{Xu2019TGRS}.
	\end{assumption}
	\begin{assumption}
	Since the RW UAV mechanical wobbling has period property, $V(t)$ is assumed periodical variation in the time domain \cite{Banagar2020TVT,Hou2019VTC}. 
	Moreover, the envelope of velocity is time-varying in practical scenarios. 
	Therefore, the value of time-varying $V(t)$ is assumed as	 
	\begin{equation}
	V\left( t \right) = a\left( t \right)\cos \left( \omega _{\mathrm{v}} t + \phi_{0}\right),
	\label{equ:Velocity}
	\tag{1}
	\end{equation}
	where $a(t)$ is the random amplitude envelope of the radial velocity under RW UAV wobbling, $\omega_{\mathrm{v}}$ is the mechanical vibration frequency from rotor, $t$ is the time, and $\phi_{0}$ is the random phase at $t=0$, which is uniformly distributed from 0 to $2\pi$. 
	\end{assumption}
	\begin{assumption}
	For the sake of simplicity, $a(t)$ is assumed to be a stationary Gaussian random process with the distribution $\mathcal{N}(0, \sigma_{\mathrm{v}}^2 )$. 
	In practical scenarios, the radial velocity of RW UAV wobbling may be influenced by wind gusts and mechanical vibration. 
	Therefore, the autocorrelation of $a(t)$ and $a\left( t + \Delta t \right)$ is assumed as $\mathrm{E}\left[a(t)a\left( t +\Delta t \right)\right]= \sigma_{\mathrm{v}}^2e^{  - \mu \Delta t  }$, where $\sigma_{\mathrm{v}}^2$ is the variance of the radial velocity and $\mu$ is the parameter to measure how fast the envelope of radial velocity changes with time.
	\end{assumption}	
	
	Based on Assumption 3, the wobbling distance $d\left( t \right)$ of the hovering RW UAV system is calculated by
	\begin{equation}
	d\left( t \right) = \int\limits_0^t {V\left( t \right)\mathrm{d}t}= \int\limits_0^t {{\mathop{\mathrm Re}\nolimits} \left\{ {a\left( t \right){e^{j{\omega _{\mathrm{v}}}t + \phi_{0} }}} \right\}\mathrm{d}t}.
	\label{equ:distance_ori}
	\tag{2}
	\end{equation}

	Based on all assumptions, patterns of UAV wobbling movement are summarized as follows:
	\begin{itemize}
		\item When $\mu\neq0$ and $\omega_{\mathrm{v}}\neq 0$, the UAV wobbling movement is influenced by random mechanical vibration and the autocorrelation of the movement velocity simultaneously.
		\item When $\mu= 0$ and $\omega_{\mathrm{v}}= 0$, the UAV moves towards to a random direction with a stationary Gaussian distributed velocity, like the FW UAV.
		\item When $\mu\neq 0$ and $\omega_{\mathrm{v}}= 0$, the UAV wobbling movement does not have mechanical vibration but moves randomly with temporal correlated velocity, i.e., wind guts. 
		\item When $\mu=0$ and $\omega_{\mathrm{v}}\neq 0$, the UAV wobbling movement is influenced by mechanical vibration with a stationary amplitude of the movement velocity.
	\end{itemize}
	
	 In this paper, we define the average wobbling movement distance of different UAV movement patterns as
	\begin{equation}
	\sigma_{\mathrm{d}}=\sqrt{\mathrm{E}\left[\left|d(t)-d(t+\Delta t)\right|^2\right]}.
	\tag{3}
	\label{equ:sigma_d}
	\end{equation}
	
	\begin{lemma}\rm
		When $\Delta t$ approaches infinity, the variance of vibration distance $\sigma_{\mathrm{d}}^2$ is computed by
		\begin{equation}
		\sigma_{\mathrm{d}}^2=
		\left\{
		\begin{array}{ll}
		\sigma_{\mathrm{v}}^2 \frac{\mu\Delta t}{\omega_{\mathrm{v}}^2+\mu^2},& \mu\neq0, \omega_{\mathrm{v}}\neq 0,\\
		\sigma_{\mathrm{v}}^2\frac{\Delta t^2}{2},& \mu= 0, \omega_{\mathrm{v}}= 0,\\
		\sigma_{\mathrm{v}}^2\frac{\Delta t}{\mu},&	\mu\neq 0, \omega_{\mathrm{v}}= 0,\\
		\sigma_{\mathrm{v}}^2\frac{1-\cos(\omega_{\mathrm{v}}\Delta t)}{\omega_{\mathrm{v}}^2},&	\mu=0, \omega_{\mathrm{v}}\neq 0.
		\end{array}
		\right.
		\label{equ:Lemma1}
		\tag{4}
		\end{equation}
	\end{lemma}

	\begin{IEEEproof}
		See Appendix A.
	\end{IEEEproof}
		
	To validate Lemma 1, $\sigma_{\mathrm{d}}^2$ are plotted against $\Delta t$ in Fig. \ref{fig:sigma_d}, where analytic results computed by \eqref{equ:Lemma1} are compared with Monte-Carlo simulation with $10^4$ realizations. 
	 Since the motor vibration frequency is high, the time separation of UAV movement caused by motor vibration will be  in millisecond scale. 
	The following observations are made therein.
	\begin{itemize}
		\item The analytical results match the simulation well. 
		\item Except for the scenario with $\mu=0$ and $\omega_{\mathrm{v}}\neq 0$, $\sigma^2_{\mathrm{d}}$ increases with $\Delta t$ enlarging. 
		\item For the scenario with $\mu=0$ and $\omega_{\mathrm{v}}\neq 0$, the UAV moves periodically.
		\item The $ \omega_{\mathrm{v}} $ has a significant impact on the amplitude of the $\sigma^2_{\mathrm{d}}$.
	\end{itemize}
	\begin{figure*}[t] 
	\begin{equation}
		\label{equ:bessel}
		\tag{7}
		\begin{array}{l}
			C\left(t, t+\Delta t\right) = C\left( \Delta t\right) = {e^{{ - 0.5\sigma_{\mathrm{v}}^2{{\left( {\frac{{{\omega _{\mathrm{c}}}}}{c\left( {\omega _{\mathrm{v}}^2 + {\mu ^2}} \right)}} \right)}^2}\left( {\mu \Delta t\left( {\omega _{\mathrm{v}}^2 + {\mu ^2}} \right) - 2\mu {\omega _{\mathrm{v}}}\sin \left( {{\omega _{\mathrm{v}}}\Delta t} \right) e^{ { - \mu \Delta t} }  +\left(  {{\mu ^2} - {\omega _{\mathrm{v}}}^2}\right) \cos \left( {{\omega _{\mathrm{v}}}\Delta t} \right)e^{{ - \mu \Delta t}}   - {\mu ^2} + {\omega _{\mathrm{v}}}^2}\right)  }}}\\
			\hspace{1.38 in}\times {J_{0}}\left( {j0.5\sigma_{\mathrm{v}}^2{{\left( {\frac{{{\omega _{\mathrm{c}}}}}{c}} \right)}^2}\frac{{ \mu \sin \left( {\omega _{\mathrm{v}}}\Delta t \right)  - {\omega _{\mathrm{v}}}\cos \left( {\omega _{\mathrm{v}}}\Delta t \right) + {\omega _{\mathrm{v}}}e^{ { - \mu \Delta t} } }}{{\left( {\omega _{\mathrm{v}}^2 + {\mu ^2}} \right){\omega _{\mathrm{v}}}}}} \right).
		\end{array}
	\end{equation}
	\hrulefill	
\end{figure*}	
	
	According to \cite{Durgin2003}, the channel impulse response (CIR) of RW UAV channel with mechanical vibration can be written as the function of the UAV wobbling distance, i.e.,
	\begin{equation}
	\label{equ:CIR}
	h\left( t \right) = h_0 {e^{j\frac{\omega _{\mathrm{c}}}{c}\int\limits_0^t {{\mathop{\mathrm Re}\nolimits} \left\{ {a\left( t \right){e^{j\left( {{\omega _{\mathrm{v}}}t + \phi_{0} } \right)}}} \right\}\mathrm{d}t} }},
	\tag{5}
	\end{equation}
	where $h_0$ is the amplitude of the CIR, $\omega_{\mathrm{c}}$ is the carrier frequency, and $c$ is speed of light. 
	
	Assuming that $ h(t) $ is stationary \cite{Cheng2019IOT}, the channel temporal ACF, defined by \cite{J.Zhang2019}, is employed to characterize the time-varying characteristics of the wireless channel. 
	Using CIR defined in \eqref{equ:CIR}, the channel temporal ACF can be computed by
	\begin{equation}
	\tag{6}
	\label{equ:ACF}
	C\left( \Delta t\right) = \frac{1}{|h_{0}|^2}\mathrm{E}\left[ {h\left( t \right){h^*}\left( {t + \Delta t} \right)} \right],
	\end{equation}
	where $(\cdot)^*$ denotes the complex conjugate operator.
	When the received signal is non-stationary, the channel temporal ACF becomes a function of both $t $ and $t + \Delta t $ and will be denoted as $ C(t, t+\Delta t) $.

	In the next section, $C\left( \Delta t\right)$ will be derived in a closed-form expression. 
	Based on the closed-form channel temporal ACF, the Doppler PSD will be calculated.

	\section{Analytical Results}
	\subsection{Closed-form Channel Temporal ACF Expression}
	The closed-form expression of the channel temporal ACF is derived in this subsection.	
	
	\begin{theorem}\rm
		The channel temporal ACF, $C\left( \Delta t \right)$, can be computed by \eqref{equ:bessel}, which is shown at the top of this page, where $J_{0}(\cdot)$ denotes Bessel function of the first kind with an order zero.
	\end{theorem}
	\begin{IEEEproof} 
		See Appendix B.
	\end{IEEEproof}	
	
	From Theorem 1, we notice that the closed-form expression of channel temporal ACF given in \eqref{equ:bessel} is directly applicable for $\mu\neq0$ and $\omega_{\mathrm{v}}\neq0$ scenario. 
	Moreover, to apply the results to the scenario with $\mu=0$ or $\omega_{\mathrm{v}}=0$, the channel temporal ACF for $(\mu\neq 0,\ \omega_{\mathrm{v}}= 0)$, $(\mu= 0,\ \omega_{\mathrm{v}}\neq 0)$, and $(\mu= 0,\ \omega_{\mathrm{v}}= 0)$ are derived as follows by using some straightforward derivations.
	
	When $\mu\neq 0,\ \omega_{\mathrm{v}}= 0$, the channel temporal ACF, $C\left( \Delta t \right)$, is computed by
	\begin{equation}
	\begin{array}{l}
	C\left( \Delta t\right) = {e^{{ - 0.5\sigma_{\mathrm{v}}^2{{\left( {\frac{{{\omega _{\mathrm{c}}}}}{c\mu}} \right)}^2}
	\left( {\mu \Delta t  + e^{{ - \mu \Delta t}}   - 1 }\right)  }}}\\
	\hspace{0.45in}\times {J_0}\left( j0.5\sigma_{\mathrm{v}}^2{{\left( {\frac{{{\omega _{\mathrm{c}}}}}{c\mu}} \right)}^2}
	\left( {\mu \Delta t  + e^{{ - \mu \Delta t}}   - 1 }\right) \right).
	\end{array}
	\tag{8}
	\label{equ:bessel_1}
	\end{equation}
	
	When $\mu= 0,\ \omega_{\mathrm{v}}\neq 0$, the channel temporal ACF, $C\left( \Delta t \right)$, is computed by
	\begin{equation}
	\begin{array}{l}
	C\left( \Delta t\right) = 
	e^{ - \frac{1}{2}\sigma_{\mathrm{v}}^2\left( \frac{\omega _{\mathrm{c}}}{c\omega _{\mathrm{v}}}\right)^2 \left(1-\cos \left( \omega _{\mathrm{v}} \Delta t \right) \right) }\\
	\hspace{0.45in} \times{J_0}\left( j\frac{1}{2}\sigma_{\mathrm{v}}^2{\left( {\frac{\omega _{\mathrm{c}}}{c\omega _{\mathrm{v}}}} \right)}^2\left(  1-\cos \left(  \omega _{\mathrm{v}} \Delta t \right) \right)  \right).
	\end{array}
	\tag{9}
	\label{equ:bessel_2}
	\end{equation}
	
	When $\mu= 0,\ \omega_{\mathrm{v}}= 0$, the channel temporal ACF, $C\left( \Delta t \right)$, is computed by
	\begin{equation}
	\tag{10}
	\label{equ:bessel_3}
	C\left( \Delta t\right) 
	=\left[e^{-\frac{\Delta t^2}{4}\sigma_{\mathrm{v}}^2{\left(\frac{{{\omega _{\mathrm{c}}}}}{c}\right)^2}}\right]
	J_0\left(j\frac{\Delta t^2}{4}\sigma_{\mathrm{v}}^2{\left(\frac{{{\omega _{\mathrm{c}}}}}{c}\right)^2}\right).
	\end{equation}
	\subsection{Doppler PSD}
	The Doppler PSDs of the proposed channel model can be computed by applying the Fourier transform of the channel temporal ACF with respect to time separation $\Delta t$ \cite{Yang2019ISAP}.
	\begin{equation}
	\label{equ:DPSD}
	\tag{11}
	F\left( {{f_D}} \right) = \int\limits_{ - \infty }^\infty  {C\left( \Delta t\right) {e^{ - j2\pi {f_{\mathrm{D}}}\Delta t}}{\mathrm{d}}\Delta t},
	\end{equation}
	where $f_{\mathrm{D}}$ is the Doppler frequency.
	
	\begin{table}
		\centering
		\caption{Simulation parameters for ACF}
		\begin{tabular}{|cc|cc|c|}
			\hline
			$\mu$&$\omega_{\mathrm{v}}$&$\sigma_{\mathrm{v}}$  [m/s]&$\sigma_{\mathrm{d}}$  [m]&Figure\\			
			\hline
			0&0&0.001${\sqrt{2}}$&0.001${\Delta t}$ &3(a)\\ 
			0&0&0.005${\sqrt{2}}$&0.005${\Delta t}$ &3(a)\\    		
			0&0&0.01${\sqrt{2}}$&0.01${\Delta t}$ &3(a)\\    
			\hline
			0&20$\pi$&0.001$\frac{\omega_{\mathrm{v}}}{\sqrt{2}}$&0.001$\sin(\frac{1}{2}\omega_{\mathrm{v}}\Delta t)$ &3(b)\\
			0&20$\pi$&0.005$\frac{\omega_{\mathrm{v}}}{\sqrt{2}}$&0.005$\sin(\frac{1}{2}\omega_{\mathrm{v}}\Delta t)$ &3(b)\\
			0&20$\pi$&0.01$\frac{\omega_{\mathrm{v}}}{\sqrt{2}}$&0.01$\sin(\frac{1}{2}\omega_{\mathrm{v}}\Delta t)$ &3(b)\\
			\hline
			30&0&0.001$\sqrt{\mu}$&0.001$\sqrt{\Delta t}$ &3(c)\\
			30&0&0.005$\sqrt{\mu}$&0.005$\sqrt{\Delta t}$ &3(c)\\
			30&0&0.01$\sqrt{\mu}$&0.01$\sqrt{\Delta t}$ &3(c)\\ 
			\hline
			30&20$\pi$&0.001$\sqrt{ \frac{\omega_{\mathrm{v}}^2+\mu^2}{\mu}}$&0.001$\sqrt{\Delta t}$&3(d)\\
			30&20$\pi$&0.005$\sqrt{ \frac{\omega_{\mathrm{v}}^2+\mu^2}{\mu}}$&0.005$\sqrt{\Delta t}$ &3(d)\\
			30&20$\pi$&0.01$\sqrt{ \frac{\omega_{\mathrm{v}}^2+\mu^2}{\mu}}$&0.01$\sqrt{\Delta t}$ &3(d)\\
			\hline
			0&10$\pi$&0.005$\frac{\omega_{\mathrm{v}}}{\sqrt{2}}$&0.005$\sin(\frac{1}{2}\omega_{\mathrm{v}}\Delta t)$ &4(a)\\
			0&20$\pi$&0.005$\frac{\omega_{\mathrm{v}}}{\sqrt{2}}$&0.005$\sin(\frac{1}{2}\omega_{\mathrm{v}}\Delta t)$ &4(a)\\
			0&30$\pi$&0.005$\frac{\omega_{\mathrm{v}}}{\sqrt{2}}$&0.005$\sin(\frac{1}{2}\omega_{\mathrm{v}}\Delta t)$ &4(a)\\
			\hline
			30&10$\pi$&0.005$\sqrt{ \frac{\omega_{\mathrm{v}}^2+\mu^2}{\mu}}$&0.005$\sqrt{\Delta t}$ &4(b)\\
			30&20$\pi$&0.005$\sqrt{ \frac{\omega_{\mathrm{v}}^2+\mu^2}{\mu}}$&0.005$\sqrt{\Delta t}$ &4(b)\\
			30&30$\pi$&0.005$\sqrt{ \frac{\omega_{\mathrm{v}}^2+\mu^2}{\mu}}$&0.005$\sqrt{\Delta t}$ &4(b)\\
			\hline
			10&0&0.005$\sqrt{\mu}$&0.005$\sqrt{\Delta t}$ &5(a)\\
			30&0&0.005$\sqrt{\mu}$&0.005$\sqrt{\Delta t}$ &5(a)\\
			50&0&0.005$\sqrt{\mu}$&0.005$\sqrt{\Delta t}$ &5(a)\\
			\hline
			10&20$\pi$&0.005$\sqrt{ \frac{\omega_{\mathrm{v}}^2+\mu^2}{\mu}}$&0.005$\sqrt{\Delta t}$ &5(b)\\
			30&20$\pi$&0.005$\sqrt{ \frac{\omega_{\mathrm{v}}^2+\mu^2}{\mu}}$&0.005$\sqrt{\Delta t}$ &5(b)\\
			50&20$\pi$&0.005$\sqrt{ \frac{\omega_{\mathrm{v}}^2+\mu^2}{\mu}}$&0.005$\sqrt{\Delta t}$ &5(b)\\
			\hline
		\end{tabular}
		\label{table:1}
	\end{table}
	\section{Numerical Results}  
	\begin{figure}[t]
		\centering
		\subfigure[$\mu =0, \sigma_{\mathrm{v}}=0.005\frac{\omega_{\mathrm{v}}}{\sqrt{2}}$]{\includegraphics[width=0.8\linewidth]{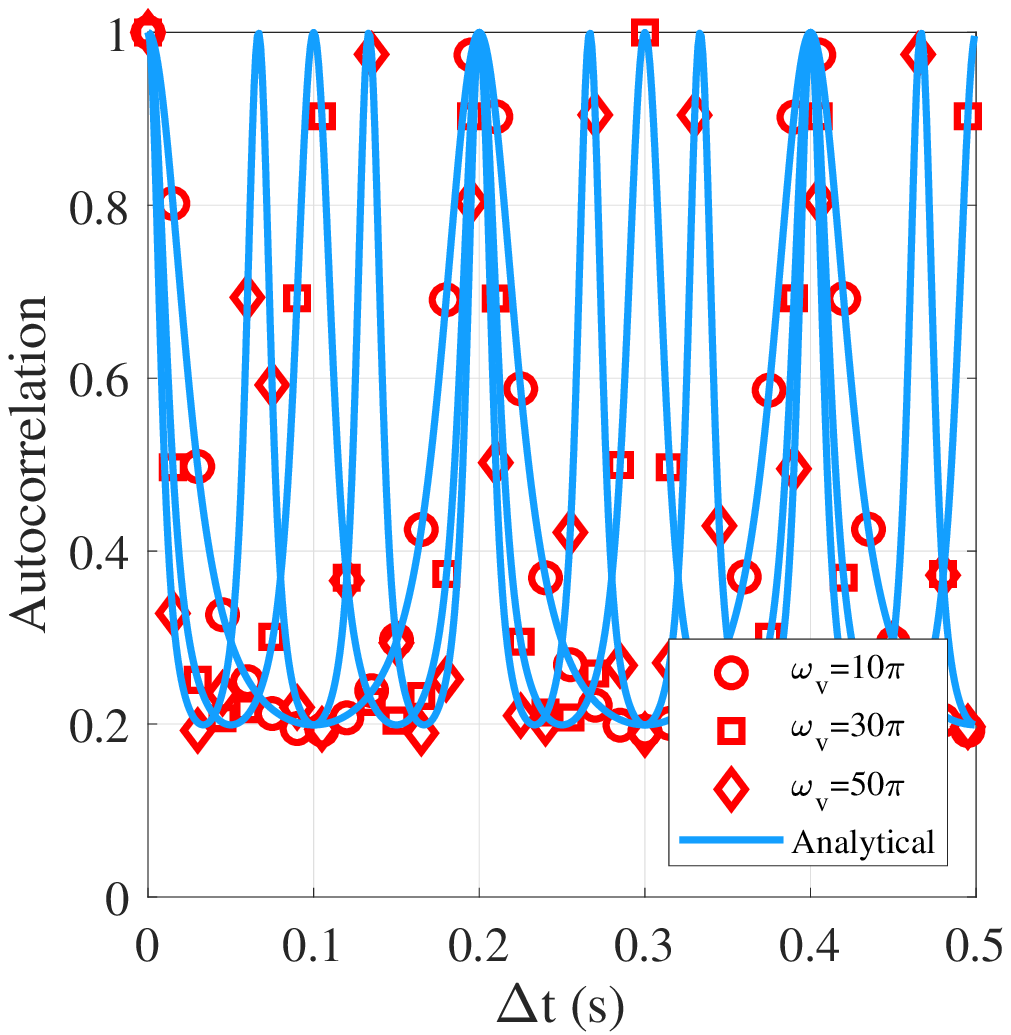}}
		\subfigure[$\mu\!=\!30,\ \sigma_{\mathrm{v}}\!=\!0.005\sqrt{\frac{\omega^2_{\mathrm{v}}\!+\!\mu^2}{\mu}}$]{\includegraphics[width=0.8\linewidth]{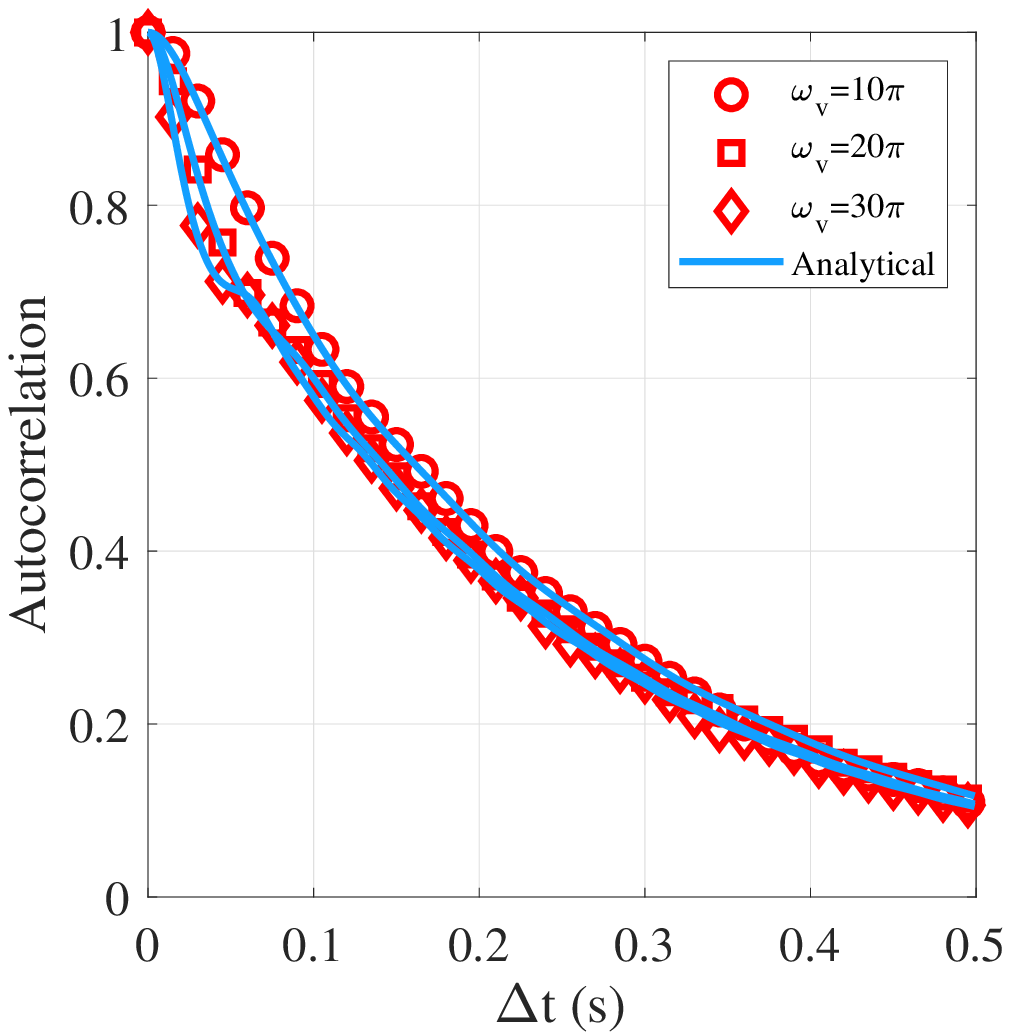}}
		\caption{The impact of $\omega_{\mathrm{v}}$ on channel temporal ACF. Solid lines illustrate analytical results and markers show simulation results.}
		\label{fig:ACF_omegav}
	\end{figure}
	
	\begin{figure}[t]
		\centering
		\subfigure[$\omega_{\mathrm{v}} =0,  \sigma_{\mathrm{v}}=0.005\sqrt{\mu}$]{\includegraphics[width=0.8\linewidth]{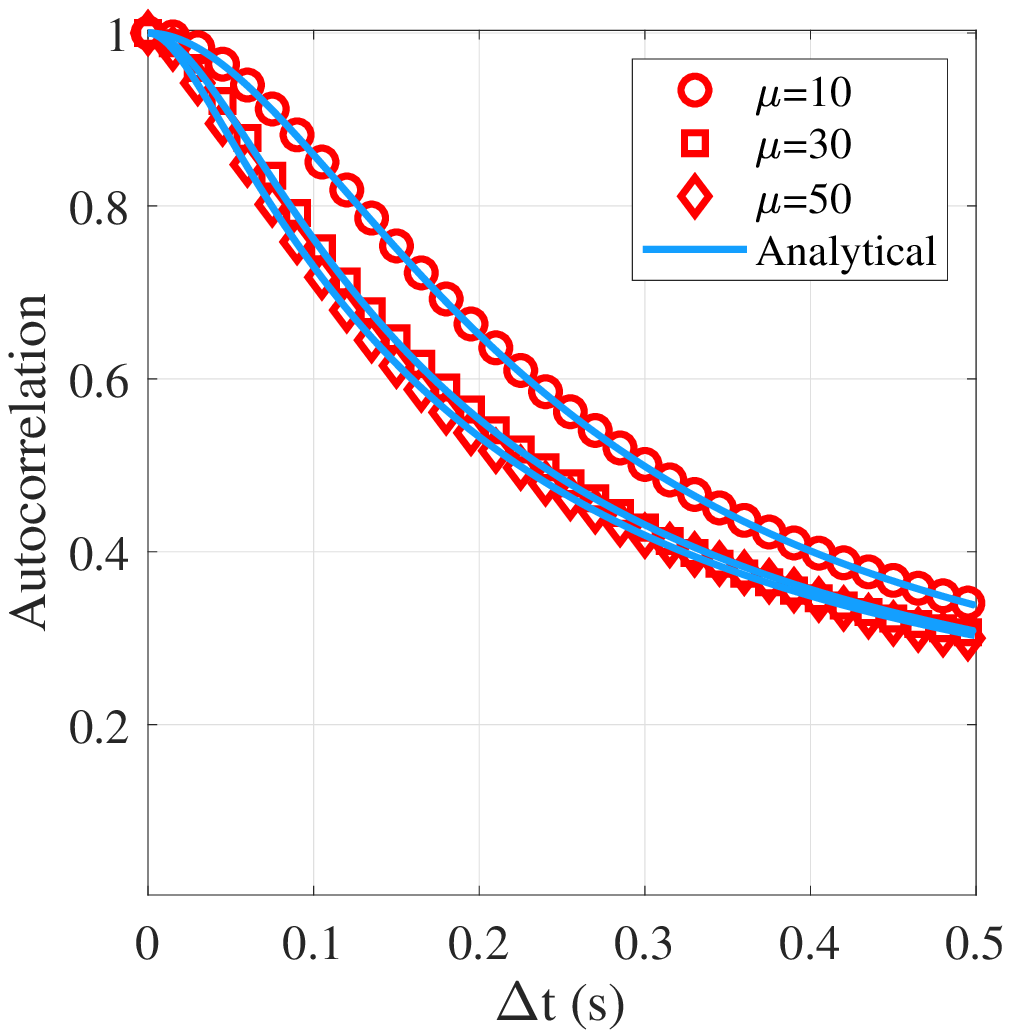}}
		\subfigure[$\omega_{\mathrm{v}}\!=\!20\pi,\  \sigma_{\mathrm{v}}\!=\!0.005\sqrt{\frac{\omega^2_{\mathrm{v}}\!+\!\mu^2}{\mu}}$]{\includegraphics[width=0.8\linewidth]{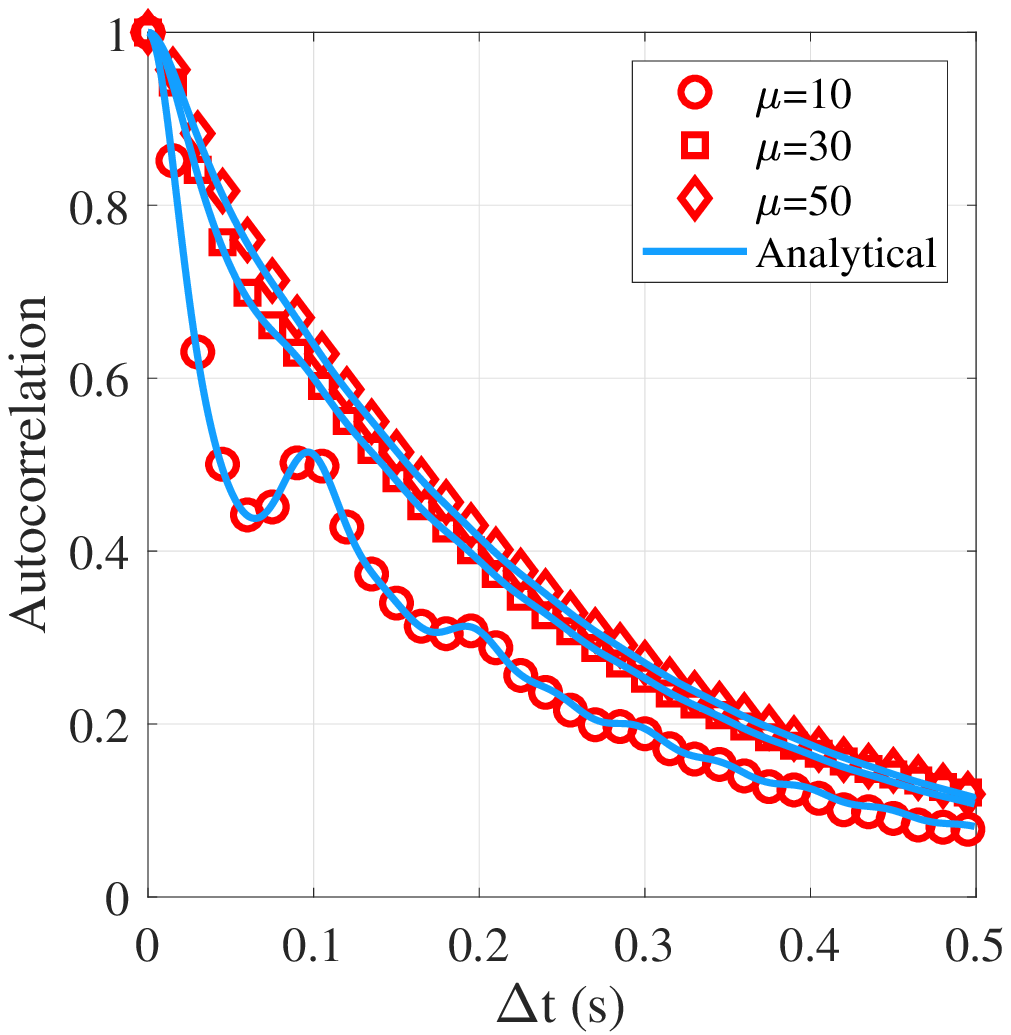}}
		\caption{The impact of $\mu$ on channel temporal ACF. Solid lines illustrate analytical results and markers show simulation results.}
		\label{fig:ACF_mu}
	\end{figure}
	
	\begin{figure}[t]
		\centering	\subfigure[$\sigma_{\mathrm{v}}=0.005\sqrt{\frac{\omega^2_{\mathrm{v}}\!+\!\mu^2}{\mu}}$]{\includegraphics[width=0.8\linewidth]{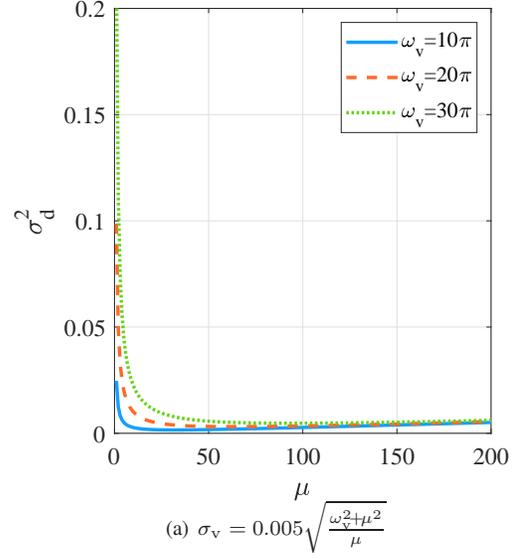}}
		\subfigure[$\omega_{\mathrm{v}}\!=\!20\pi,\  \sigma_{\mathrm{v}}\!=\!0.005\sqrt{\frac{\omega^2_{\mathrm{v}}\!+\!\mu^2}{\mu}}$]{\includegraphics[width=0.8\linewidth]{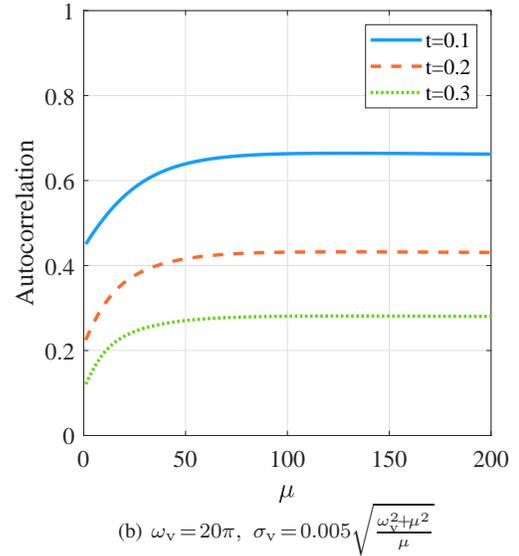}}
		\caption{ (a) The impact of $\mu$ on $\sigma_{\mathrm{d}}^2$ with different $\omega_{\mathrm{v}}$; (b) The impact of $\mu$ on channel temporal ACF at different time slot.}
		\label{fig:proof_mu}
	\end{figure}
\begin{figure}[t]
	\centering
	\subfigure[$\omega_{\mathrm{v}} =0,  \mu=0$]{\includegraphics[width=0.8\linewidth]{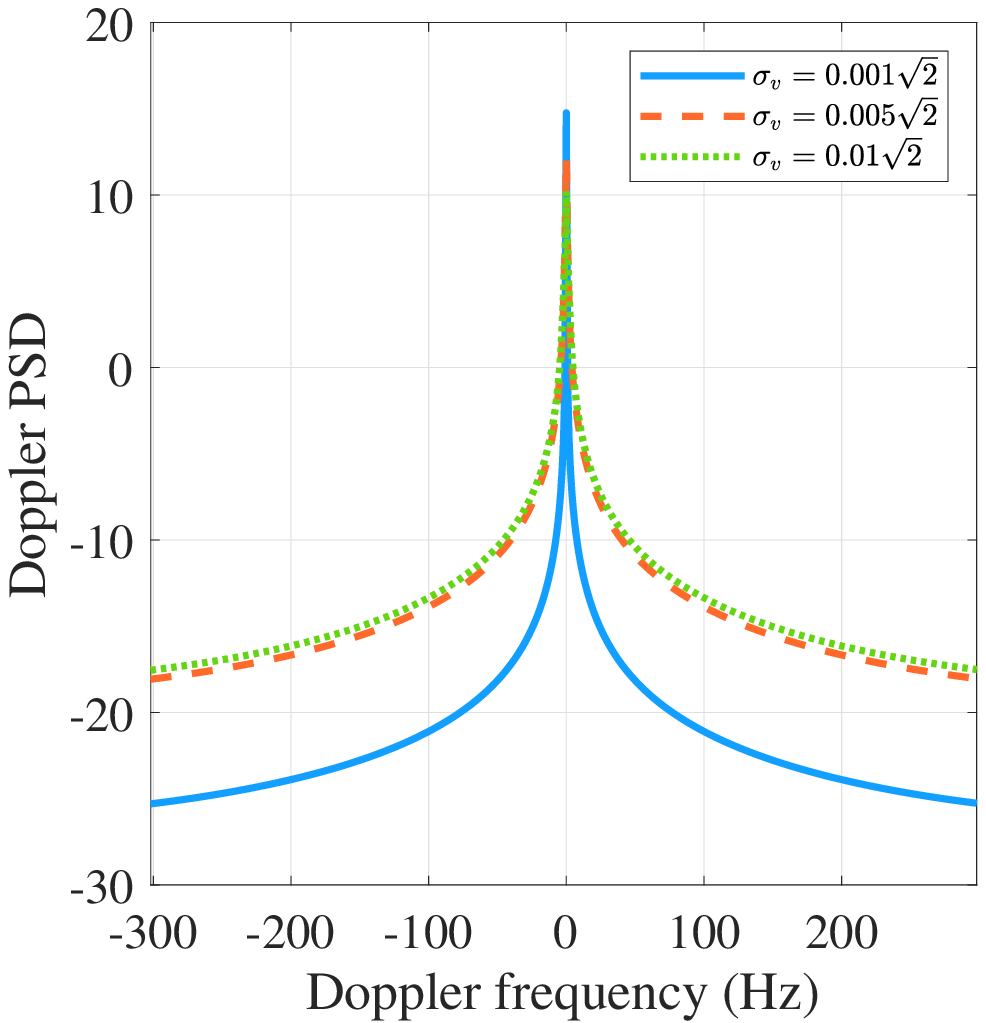}}
	\subfigure[$\omega_{\mathrm{v}} =200\pi, \mu=30$]{\includegraphics[width=0.8\linewidth]{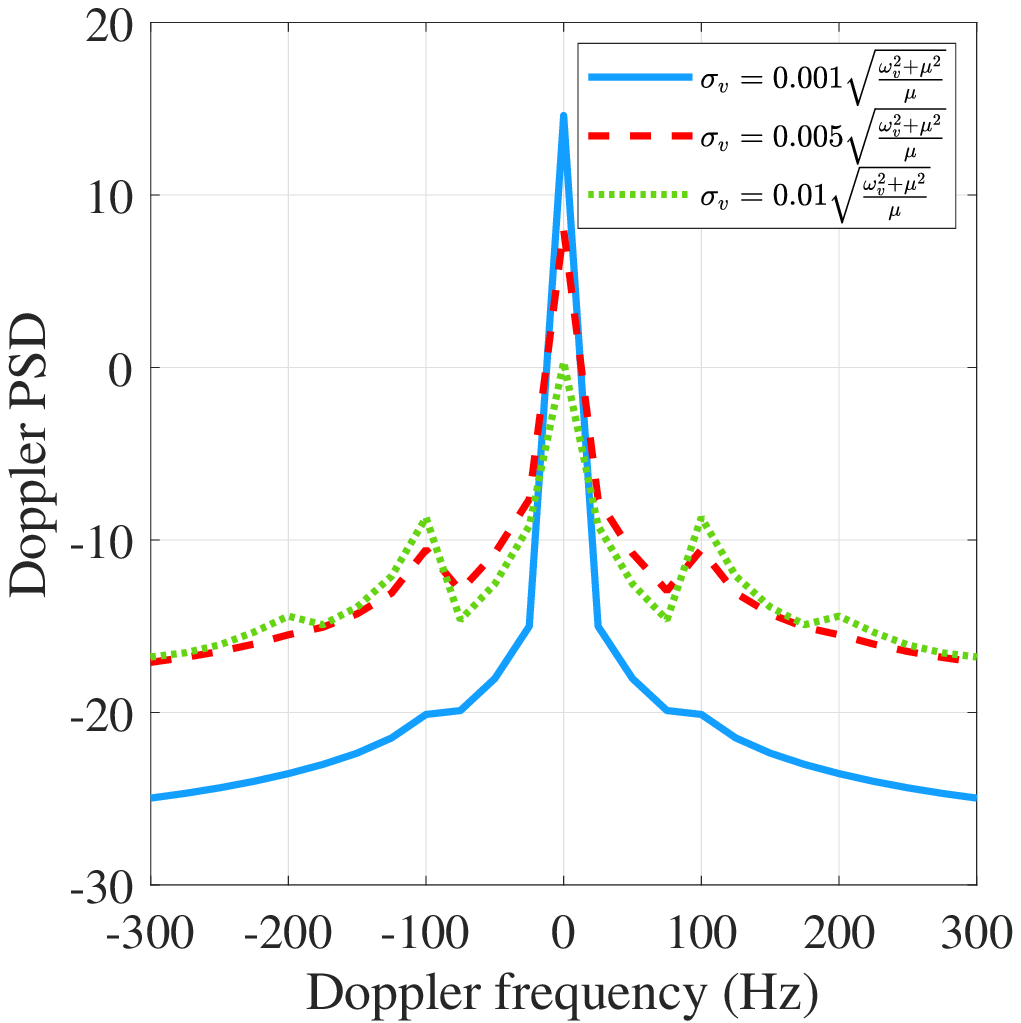}}
	\caption{The impact of $ \sigma_{\mathrm{v}} $ on Doppler PSD with different $\mu$ and $\omega_{\mathrm{v}}$.}
	\label{fig:Doppler_sigmad}
\end{figure}
\begin{figure}[t]
	\centering
	\includegraphics[width=0.8\linewidth]{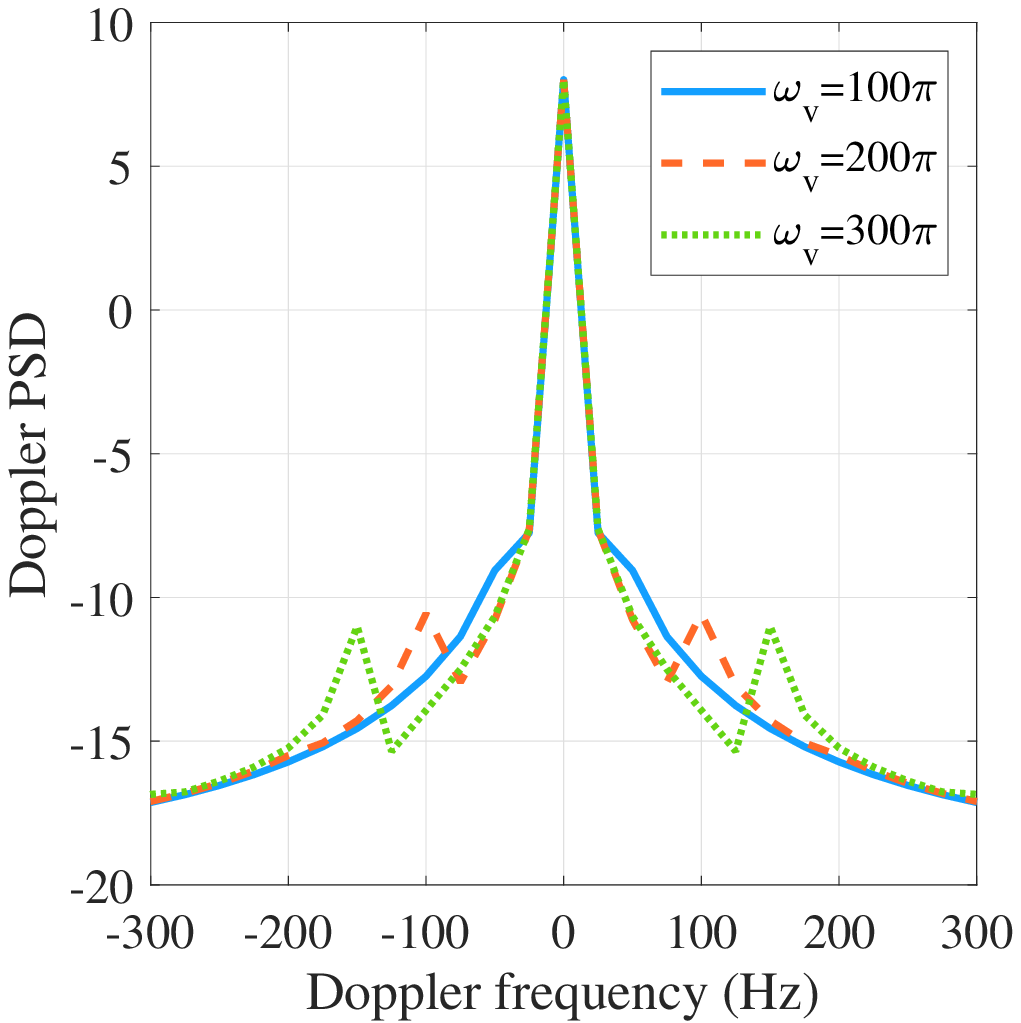}
	\caption{The impact of $\omega_{\mathrm{v}}$ on Doppler PSD with $ \sigma_{\mathrm{v}}=0.005\sqrt{\frac{\omega^2_{\mathrm{v}}+\mu^2}{\mu}} $ and $ \mu=30 $. }
	\label{fig:Doppler_omegav}
	\centering
	\includegraphics[width=0.8\linewidth]{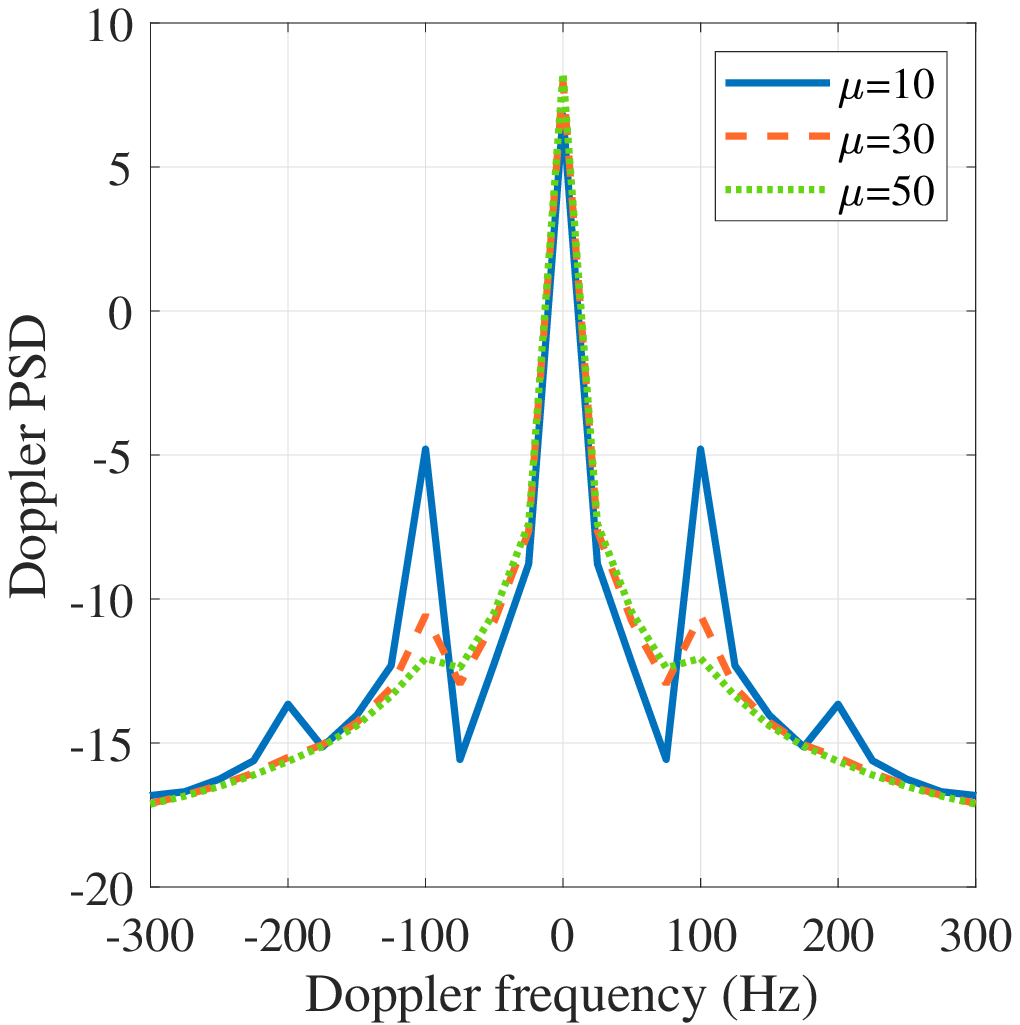}
	\caption{The impact of $\mu$ on Doppler PSD with $\sigma_{\mathrm{v}}=0.005\sqrt{\frac{\omega^2_{\mathrm{v}}+\mu^2}{\mu}}$ and $\omega_{\mathrm{v}}=200\pi$. }
	\label{fig:Doppler_mu}
\end{figure}
\begin{figure*}[t]
	\normalsize
	\begin{eqnarray}
		\setcounter{equation}{12}
		\label{equ:lemma1_1}
		&&\sigma_{\mathrm{d}}^2={\mathrm{E}}\left[ {\int\limits_0^{\Delta t} {a\left( {{t_1}} \right)\cos \left( {{\omega _{\mathrm{v}}}{t_1} + {\phi _0}} \right){\rm{d}}{t_1}} \int\limits_0^{\Delta t} {a\left( {{t_2}} \right)\cos \left( {{\omega _{\mathrm{v}}}{t_2} + {\phi _0}} \right){\rm{d}}{t_2}} } \right] \nonumber\\
		&&\hspace{0.2in}= {\mathrm{E}}\left[ {\int\limits_0^{\Delta t} {\int\limits_0^{\Delta t} {a\left( {{t_1}} \right)a\left( {{t_2}} \right)\cos \left( {{\omega _{\mathrm{v}}}{t_1} + {\phi _0}} \right)\cos \left( {{\omega _{\mathrm{v}}}{t_2} + {\phi _0}} \right){\rm{d}}{t_1}{\rm{d}}{t_2}} } } \right].
	\end{eqnarray}
	\hrulefill
				\begin{equation}
		\tag{16}
		\label{equ:lemma1_final}
		\sigma_{\mathrm{d}}^2= \sigma_{\mathrm{v}}^2\frac{{\mu \Delta t}}{{{\mu ^2} + \omega _{\mathrm{v}}^2}} +\sigma_{\mathrm{v}}^2 \frac{{\omega _{\mathrm{v}}^2 - {\mu ^2} - \omega _{\mathrm{v}}^2{e^{ - \mu \Delta t}}\cos \left( {{\omega _{\mathrm{v}}}\Delta t} \right) - 2\mu {\omega _{\mathrm{v}}}{e^{ - \mu \Delta t}}\sin \left( {{\omega _{\mathrm{v}}}\Delta t} \right) + {\mu ^2}{e^{ - \mu \Delta t}}\cos \left( {{\omega _{\mathrm{v}}}\Delta t} \right)}}{{{{\left( {{\mu ^2} + \omega _{\mathrm{v}}^2} \right)}^2}}}.
	\end{equation}
	\hrulefill
\end{figure*}	

	\begin{figure}[t]
	\centering
	\subfigure[PSK]{\includegraphics[width=0.8\linewidth]{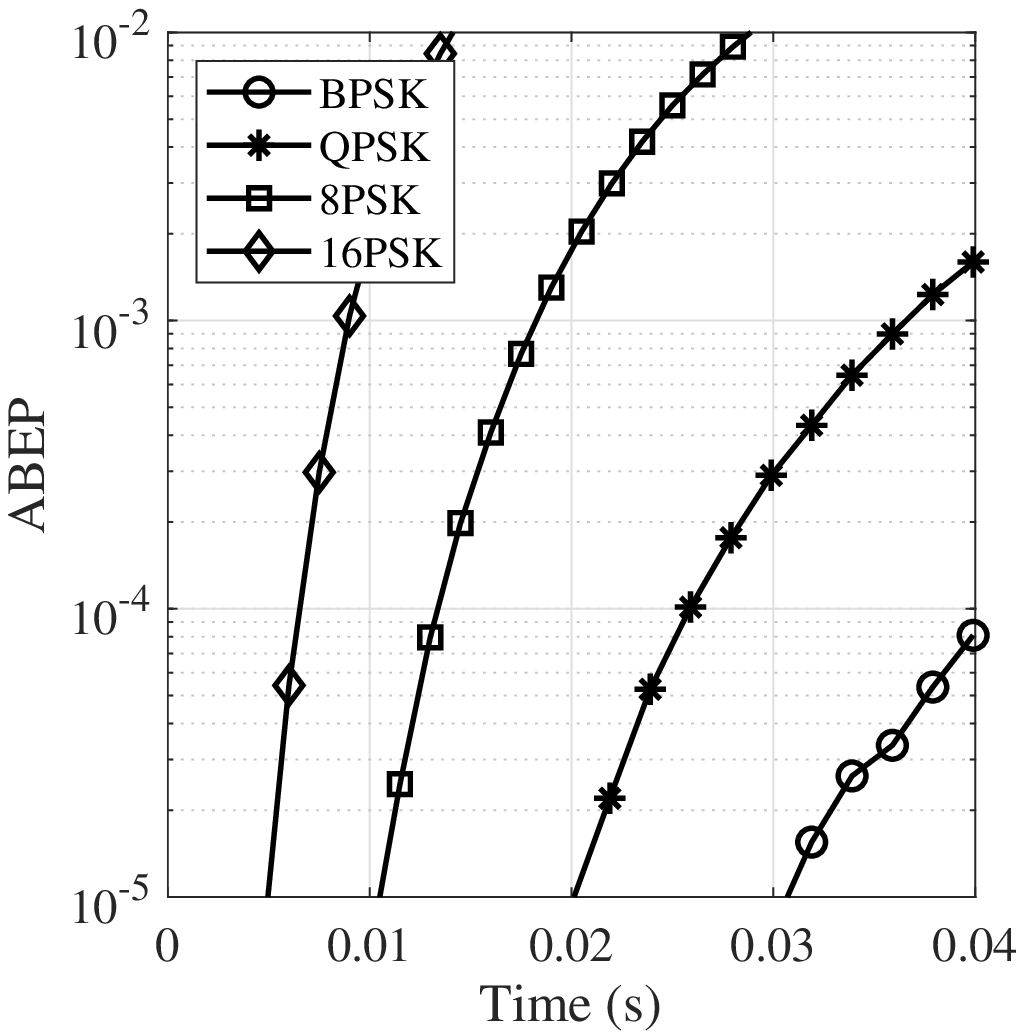}}
	\subfigure[QAM]{\includegraphics[width=0.8\linewidth]{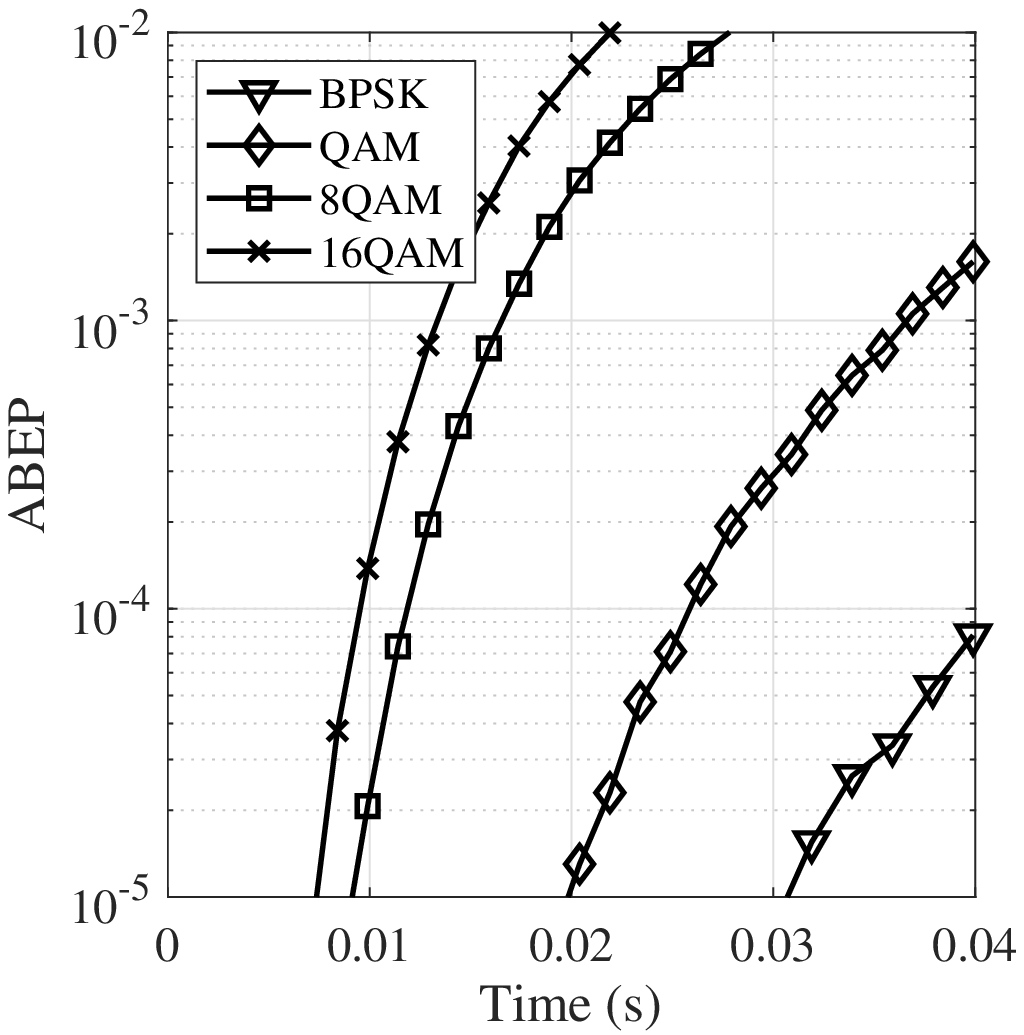}}
	\caption{The ABEP performance of millimeter-wave RW UAV A2G link under mechanical wobbling employing with PSK and QAM signals.}
	\label{fig:MISO_Time_ABEP_1E5}
\end{figure}
	In this section, the impacts of the mechanical vibration frequency and the velocity envelope covariance on channel temporal ACF and Doppler PSD of millimeter-wave RW UAV channel are investigated.
	
	The numerical results of channel temporal ACF are provided. 
	The parameters used for channel temporal ACF simulations are summarized in Table \ref{table:1}. 
	 In the simulation, the carrier frequency is 28 GHz, which is widely used in the millimeter-wave UAV channel \cite{Khawaja2017VTC,Michailidis2019TVT}.
	Since the system is working in the millimeter-wave bands and the mechanical wobbling is millimeter-wave-scale, the distance variance parameters could be set as the 1 mm, 5 mm, and 10 mm, which is presented in $\sigma_{\mathrm{v}}$ and $\sigma_{\mathrm{d}}$ as the constant 0.001, 0.005, and 0.01, respectively.
	When $\mu=10,\ 30$, and $50$ at $\Delta t=0.001 $ second, it means the $\mathrm{E}\left[a(t)a\left( t +\Delta t \right)\right]= 0.99,\ 0.97$, and 0.95, respectively. 
	If the autocorrelation of $ a(t)$ and $a\left(t+\Delta t \right) $ is low, the channel temporal ACF will decrease too fast.
	When $\omega_{\mathrm{v}}=10\pi,\ 20\pi$, and $30\pi$, it means that mechanical vibration change frequencies are 5 Hz, 10 Hz, and 15 Hz, respectively. 

	Fig. \ref{fig:ACF_sigmad} shows the channel temporal ACFs for analytical and simulation model at different $\sigma_{\mathrm{v}}$. 
	The simulation and analytical results fit well, which ensures the correctness of our derivations.
	According to Fig. \ref{fig:ACF_sigmad}, when $\sigma_{\mathrm{v}}$ increases, the channel temporal ACF decreases faster for all cases. 
	In different scenarios cases, the channel temporal ACF presents different models. 
	For Fig. \ref{fig:ACF_sigmad} (b), since $\mu=0$ and $\omega_{\mathrm{v}}\neq 0$, the UAV wobbling movement is influenced by mechanical vibration with a stationary amplitude of the movement velocity. 
	Hence, the channel temporal ACF changes with mechanical vibration in periodic.
	When $\mu\neq0$ and $\omega_{\mathrm{v}}\neq 0$, the $\sigma_{\mathrm{v}}$ has a more significant impact of channel temporal ACF than that of other scenarios.
	
	Fig. \ref{fig:ACF_omegav} presents the impact of $\omega_{\mathrm{v}}$ on channel temporal ACF. 
	When $\omega_{\mathrm{v}}$ enhances, according to Fig. \ref{fig:ACF_omegav} (a), the channel temporal ACF changing frequency is increase. 
	Since $\omega_{\mathrm{v}}$ enhances, it also caused $\sigma_{\mathrm{v}}$ boost, which results in the decrease rate of channel temporal ACF rising.
	According to Fig. \ref{fig:ACF_omegav} (b), the channel temporal ACF has some fluctuations at the value of $\omega_{\mathrm{v}}$ much bigger than $\mu$.
	Meanwhile, the simulation results are compared with the analytical results to ensure the correctness of our derivations.
	
	Fig. \ref{fig:ACF_mu} illustrates the impact of $\mu$ on channel temporal ACF. 
	For Fig. \ref{fig:ACF_mu} (a), the channel temporal ACF decreases faster, when $\mu$ increases.
	In this situation, the UAV wobbling movement does not have mechanical vibration. 
	Therefore, the velocity change may be caused by environmental factors, i.e., wind gusts. 
	For Fig. \ref{fig:ACF_mu} (b), when $\mu$ increases with mechanical vibration, the channel temporal ACF will decrease slower. 
	The reason for this situation is $\mu$ has a greater effect of the  increase or decrease trend of $\sigma_{\mathrm{d}}$ than $\omega_{\mathrm{v}}$ at $\mu\neq 0$ and $\omega_{\mathrm{v}}\neq0$ scenario.
	
	Fig. \ref{fig:proof_mu} can be used to explain why the decrease rate of channel temporal ACF abate when $\mu$ rise in Fig. \ref{fig:ACF_mu} (b).
	According to Fig. \ref{fig:proof_mu} (a), when $\mu$ increases, $\sigma_{\mathrm{d}}^2$ will decrease firstly and then increase very slow. 
	According to Fig. \ref{fig:ACF_sigmad}, when $\sigma_{\mathrm{v}}$ increases, the decrease rate of channel temporal ACF will boost.
	Therefore, the decrease rate of channel temporal ACF will abate when $\mu$ increases at $\mu\neq 0$ and $\omega_{\mathrm{v}}\neq0$ scenarios.
	Moreover, in Fig. \ref{fig:proof_mu} (b), at the same time slot, when $\mu$ enhances, the channel temporal ACF enhance firstly and then keep flat because the value of $\sigma_{\mathrm{d}}^2$ change little.
	
	Fig. \ref{fig:Doppler_sigmad} shows the impact of $\sigma_{\mathrm{v}}$ on Doppler PSD. 
	We can observe that the shape of the Doppler PSD slightly widens along the Doppler frequencies axis.
	According to Fig. \ref{fig:Doppler_sigmad} (a), when $\mu$ and $\omega_{\mathrm{v}}$ are all 0, there are no Doppler PSD drift over time.
    It means when the UAV is moving toward the ground in fix velocity, like the FW UAV, the Doppler PSD is the steepest.
	Furthermore, the value of Doppler PSD will increase, when $\sigma_{\mathrm{v}}$ enhances. 
	For Fig. \ref{fig:Doppler_sigmad} (b), it is observed that the Doppler PSD drift due to the RW UAV wobbling movement. 
	Moreover, as $\sigma_{\mathrm{v}}$ increasing, the fluctuation of Doppler PSD will be widened. 
	
	Fig. \ref{fig:Doppler_omegav} compares the different value of $\omega_{\mathrm{v}}$ on Doppler PSD. 
	When the value of $\omega_{\mathrm{v}}$ rises, the fluctuation range of Doppler PSD broadens.
	The Doppler frequency of the bulge is rise when the value of $\omega_{\mathrm{v}}$ increases.
	The fluctuation becomes significant as $\omega_{\mathrm{v}}$ enhances. 
	
	Fig. \ref{fig:Doppler_mu} illustrates the impact of $\mu$ on Doppler PSD. 
	When $\mu$ increases, the fluctuation of Doppler PSD abates. 
	However, $\mu$ will not change the Doppler frequency of the bulge. 
	Moreover, the value of $\mu$ influences the number of Doppler PSD bulges.
	
	Fig. \ref{fig:MISO_Time_ABEP_1E5} shows the average bit error probability (ABEP) performance of millimeter-wave UAV A2G link employing with \textit{M}-ary PSK or \textit{M}-ary QAM signals under mechanical wobbling. 
	The $\mu $ is 30 and  $\omega_{\mathrm{v}} $ is $ 20 \pi $ in this case.
	For millimeter-wave UAV A2G link, the channel estimation process finished at $ t=0 $ and then the signal transmission begins. 
	According to \ref{fig:MISO_Time_ABEP_1E5}, the ABEP performance of the transmission becomes worse with time under mechanical wobbling.
	A key observation is that even for small UAV wobbling, the ABEP of the millimeter-wave RW UAV A2G link deteriorates quickly, which may lead to a unreliable communication link.
	\section{Conclusions and Future Works}
	In this paper, we have proposed the analytical model of the Doppler effect brought by the mechanical wobbling in the millimeter-wave RW UAV A2G link. 
	Applying the RW UAV wobbling movement model at hovering status, the closed-form expression of channel temporal ACF has been derived and verified via Monte-Carlo simulation.
	Moreover, the Doppler PSD of millimeter-wave RW UAV channel has been computed based on the analytical channel temporal ACF. 
	Numerical results show that the mechanical vibration frequency and the radial velocity envelope covariance interact on the decrease rate and fluctuation model of channel temporal ACF.
	The Doppler spread range broadens as mechanical vibration frequency enhances while the value and number of Doppler PSD bulge degrade as radial velocity envelope covariance increases.
	A key observation is that even for small UAV wobbling, the BEP of the UAV A2G link deteriorates quickly, making the link difficult to establish a reliable communication link.
	Therefore, the UAV designer has to jointly consider the vibration frequency and the radial velocity envelope covariance carefully to mitigate the impact of the Doppler effect brought by the RW UAV mechanical wobbling on the A2G wireless link.
	
	In the future, the following issues need to be addressed.
	(1) The low-altitude RW UAV scenario will be considered because the probability of the LoS link will decrease and the tangential velocity will lead to the Doppler effect when UAV located at low-altitude \cite{Chang2019ACCESS} \cite{Zeng2019PIEEE}.
	(2) The mechanical vibration frequency and the radial velocity envelope covariance, i.e., $ \omega_{\mathrm{v}}$ and $\mu $ will be measured practically. 
	(3) Beyond the A2G wireless channel, the air-to-air (A2A) wireless channel will be modeled considering the UAV wobbling \cite{Zeng2019PIEEE}. 
	(4) How the wobbling of RW UAV wireless channel impacts the performance of UAV small cell networks will be evaluated. 
	The coverage rate and data rate will be evaluated for various system schemes \cite{Zeng2019PIEEE} \cite{Ai-hourani2014WCL}.
	\appendices
	\section*{Appendix}
	\subsection{Proof of Lemma 1}
	Substituting \eqref{equ:distance_ori} into \eqref{equ:sigma_d}, the $\sigma_{\mathrm{d}}^2$ with different time slots $ t_1 $ and $ t_2 $ can be expressed by \eqref{equ:lemma1_1}, which is shown at the top of this page. 
	
	Since $\mathrm{E}\left[a(t)a\left( t +   \Delta t \right)\right]= \sigma_{\mathrm{v}}^2e^{-\mu \Delta t}$, the ${\mathrm{E}}\left[ {a\left( {{t_1}} \right)a\left( {{t_2}} \right)} \right]$ can be expressed by
	\begin{equation}
	\label{equ:lemma1_a}
	\tag{13}
	{\mathrm{E}}\left[ {a\left( {{t_1}} \right)a\left( {{t_2}} \right)} \right] = \sigma_{\mathrm{v}}^2{e^{ - \mu \left| {{t_1} - {t_2}} \right|}}.
	\end{equation}
	
	Moreover, the ${\mathrm{E}}\left[ {\cos \left( {{\omega _{\mathrm{v}}}{t_1} + {\phi _0}} \right)\cos \left( {{\omega _{\mathrm{v}}}{t_2} + {\phi _0}} \right)} \right] $ can be calculated by
	\begin{equation}
	\label{equ:lemma1_omega}
	\tag{14}
	\begin{array}{l}
	\hspace{-0.15in}{\mathrm{E}}\left[ {\cos \left( {{\omega _{\mathrm{v}}}{t_1} + {\phi _0}} \right)\cos \left( {{\omega _{\mathrm{v}}}{t_2} + {\phi _0}} \right)} \right]\\
	= 0.5{\rm{E}}\left[ {\cos \left( {2\phi_0  + {\omega _{\mathrm{v}}}{t_1} + {\omega _{\mathrm{v}}}{t_2}} \right) + \cos \left( {{\omega _{\mathrm{v}}}\left| {{t_1} - {t_2}} \right|} \right)} \right]\\
	= 0.5\cos \left( {{\omega _{\mathrm{v}}}\left| {{t_1} - {t_2}} \right|} \right).
	\end{array}
	\end{equation}
	
	Then, substitute \eqref{equ:lemma1_a} and \eqref{equ:lemma1_omega} into \eqref{equ:lemma1_1}, the $\sigma_{\mathrm{d}}^2$ could be shown as
	\begin{equation}
	\tag{15}
	\label{equ:lemma1_2}
	\begin{array}{l}
	\sigma_{\mathrm{d}}^2 = 0.5\sigma_{\mathrm{v}}^2\int\limits_0^{\Delta t} {\int\limits_0^{{t_2}} {{e^{ - \mu \left( {{t_2} - {t_1}} \right)}}\cos \left( {{\omega _{\mathrm{v}}}\left( {{t_2} - {t_1}} \right)} \right){\rm{d}}{t_1}{\rm{d}}{t_2}} } \\
	\hspace{0.2in}+ 0.5\sigma_{\mathrm{v}}^2\int\limits_0^{\Delta t} {\int\limits_{{t_2}}^{\Delta t} {{e^{ - \mu \left( {{t_1} - {t_2}} \right)}}\cos \left( {{\omega _{\mathrm{v}}}\left( {{t_1} - {t_2}} \right)} \right){\rm{d}}{t_1}{\rm{d}}{t_2}} }. 
	\end{array}
	\end{equation} 
	
	After integral computation of \eqref{equ:lemma1_2}, the result is shown in \eqref{equ:lemma1_final}, which is shown at the top of this page. 
	
	The $ \sigma_{\mathrm{v}}^2\frac{{\mu \Delta t}}{{{\mu ^2} + \omega _{\mathrm{v}}^2}}$ part in \eqref{equ:lemma1_final} shows the main increase trend of the $\sigma_{\mathrm{d}}^2$ when $\Delta t$ approaches infinity.
	When $\mu\neq0$, $\omega_{\mathrm{v}}\neq 0$ and $\mu\neq0$, $\omega_{\mathrm{v}}=0$, the result shown in \eqref{equ:Lemma1} only considers the $ \sigma_{\mathrm{v}}^2\frac{{\mu \Delta t}}{{{\mu ^2} + \omega _{\mathrm{v}}^2}}$ part in \eqref{equ:lemma1_final}, because this part have the main influence of the value $\sigma_{\mathrm{d}}^2$.
	When $\omega_{\mathrm{v}}= 0$, $\mu=0$ and $\omega_{\mathrm{v}}\neq 0$, $\mu=0$, the result shown in \eqref{equ:Lemma1} can be computed by direct mathematical computation based on  \eqref{equ:lemma1_final}. 
	\subsection{Proof of Theorem 1}
	\begin{figure*}[t]
		\normalsize
		\begin{equation}
			\label{equ:expression}
			\tag{20}
			C\left( t, t+\Delta t\right)  =E\left[ e^{ - j\frac{\omega _{\mathrm{c}}}{c}\mathop {\lim }\limits_{N \to \infty } \left\{ {\frac{\Delta t}{N}\sum\limits_{n = 0}^{N - 1} {\left[ {\left( {{e^{ - \mu \frac{{n\Delta t}}{N}}}{b_1} + \sqrt {1 - {e^{ - \mu \frac{{2\Delta t}}{N}}}} \mathop \sum \limits_{k = 2}^{n + 1} {e^{ - \mu \frac{{(n + 1 - k)\Delta t}}{N}}}{b_k}} \right)\cos \left( {{\omega _{\mathrm{v}}}\left( {t + \frac{{n\Delta t}}{N}} \right) + \phi_0 } \right)} \right]} } \right\}} \right].
		\end{equation}
		\hrulefill
			\begin{equation}   	
			\label{equ:A}
			\tag{23}
			\alpha = \mathop {\lim }\limits_{N \to \infty } \left\{ {{{\left( {\frac{{\Delta t}}{N}} \right)}^2}{{\left( {\sum\limits_{n = 0}^{N - 1} {\cos \left( {{\omega _{\mathrm{v}}}t + {\omega _{\mathrm{v}}}\frac{{n\Delta t}}{N} + {\phi _0}} \right)e^{ { - \mu \Delta t} }} } \right)}^2}} \right\},
		\end{equation}
		
		\begin{equation}
			\label{equ:B}
			\tag{24}
			\beta = \mathop {\lim }\limits_{N \to \infty } \left\{ {{{\left( {\frac{{\Delta t}}{N}} \right)}^2}\left( {1 - e^{ { - \frac{{2\mu\Delta t}}{N}}} } \right)\sum\limits_{k = 2}^N {{{\left( {\mathop \sum \limits_{n = k - 1}^{N - 1} \cos \left( {{\omega _{\mathrm{v}}}t + {\omega _{\mathrm{v}}}\frac{{n\Delta t}}{N} + {\phi _0}} \right) e^{{ - \mu \frac{{\left( {n + 1 - k} \right)\Delta t}}{N}} } } \right)}^2}} } \right\}.
		\end{equation}
	\begin{equation}
		\label{equ:A_int}
		\tag{25}
		\alpha = {\left( {\int_0^{\Delta t} {\cos \left( {{\omega _{\mathrm{v}}}x + {\omega _{\mathrm{v}}}t + \phi_{0} } \right)e^{ { - \mu \Delta t} }{\mathrm{d}}x} } \right)^2},
	\end{equation}
	
	\begin{equation}
		\label{equ:B_int}
		\tag{26}
		\beta =2\mu \int_0^{\Delta t} {{{\left( {\int_{{x_2}}^{\Delta t} {\cos \left( {{\omega _{\mathrm{v}}}t +{\omega _{\mathrm{v}}} {x_1} + \phi_{0} } \right) e^ { - \mu {x_1} + \mu {x_2}}} } \mathrm{d}{x_1}\right) }^2}\mathrm{d}{x_2}}.
	\end{equation}
\begin{equation}
	\label{equ:key}
	\tag{27}
	\sigma _{X}^2\left( {\Delta t} \right)  = \left\{ {{{\left( {\frac{{{\omega _{\mathrm{c}}}}}{c}} \right)}^2}\left[ \begin{array}{l}
			\frac{1}{{2\left( {{\mu ^2} + {\omega _{\mathrm{v}}}^2} \right){\omega _{\mathrm{v}}}}}\left[ \begin{array}{l}
				- {\omega _{\mathrm{v}}}\cos \left( {2\phi } \right) - \mu \sin \left( {2\phi } \right) - {\omega _{\mathrm{v}}}\cos \left( {2\phi  + 2{\omega _{\mathrm{v}}}\Delta t} \right)\\
				+ \mu \sin \left( {2{\omega _{\mathrm{v}}}\Delta t + 2\phi } \right) + 2{\omega _{\mathrm{v}}}e^{ { - \mu \Delta t} }\cos \left( {{\omega _{\mathrm{v}}}\Delta t + 2\phi } \right)
			\end{array} \right]\\
			+ {\left( {\frac{1}{{{\mu ^2} + {\omega _{\mathrm{v}}}^2}}} \right)^2}\left[ \begin{array}{l}
				\mu \Delta t\left( {\omega _{\mathrm{v}}^2 + {\mu ^2}} \right) - 2\mu {\omega _{\mathrm{v}}}\sin \left( {{\omega _{\mathrm{v}}}\Delta t} \right) e^{ { - \mu \Delta t} }  - {\mu ^2} + {\omega _{\mathrm{v}}}^2\\
				+ {\mu ^2}\cos \left( {{\omega _{\mathrm{v}}}\Delta t} \right)e^{ { - \mu \Delta t} }  - {\omega _{\mathrm{v}}}^2\cos \left( {{\omega _{\mathrm{v}}}\Delta t} \right) e^{ { - \mu \Delta t} }
			\end{array} \right]
		\end{array} \right]} \right\}.
\end{equation}

\begin{equation}
	\label{equ:End}
	\tag{28}
	\sigma _{X}^2\left( {\Delta t} \right) = \left\{ {{{\left( {\frac{{{\omega _{\mathrm{c}}}}}{c}} \right)}^2}\left[ \begin{array}{l}
			\frac{1}{{\left( {{\mu ^2} + {\omega _{\mathrm{v}}}^2} \right){\omega _{\mathrm{v}}}}}\left( { - {\omega _{\mathrm{v}}}\cos \left( {\omega _{\mathrm{v}}}\Delta t\right)  + \mu \sin \left( {\omega _{\mathrm{v}}}\Delta t \right) + {\omega _{\mathrm{v}}}e^{ { - \mu \Delta t} }} \right)\cos \phi '\\
			+ {\left( {\frac{1}{{{\mu ^2} + {\omega _{\mathrm{v}}}^2}}} \right)^2}\left[ \begin{array}{l}
				\mu \Delta t\left( {\omega _{\mathrm{v}}^2 + {\mu ^2}} \right) - 2\mu {\omega _{\mathrm{v}}}\sin \left( {{\omega _{\mathrm{v}}}\Delta t} \right)e^{ { - \mu \Delta t} } - {\mu ^2} + {\omega _{\mathrm{v}}}^2\\
				+ {\mu ^2}\cos \left( {{\omega _{\mathrm{v}}}\Delta t} \right) e^{  - \mu \Delta t } - {\omega _{\mathrm{v}}}^2\cos \left( {{\omega _{\mathrm{v}}}\Delta t} \right) e^{ { - \mu \Delta t} } 
			\end{array} \right]
		\end{array} \right]} \right\}.
\end{equation}
	
			\hrulefill
	\end{figure*}
	Based on the CIRs of millimeter-wave UAV A2G link shown in \eqref{equ:CIR}, the channel temporal ACF (non-stationary case) can be computed by 
	\begin{equation}
	\tag{17}
	\label{equ:ACF_int}
	C\left( t, t + \Delta t\right) = \mathrm{E}\left[ {{e^{ - j\frac{{{\omega _{\mathrm{c}}}}}{c}{\mathop{\mathrm Re}\nolimits} \left\{ {\int\limits_t^{t + \Delta t} {a\left( t \right){e^{j\left( {{\omega _{\mathrm{v}}}t + \phi_{0} } \right)}}\mathrm{d}t} } \right\}}}} \right].
	\end{equation}
	
	It can be seen that the $C\left( t, t+\Delta t\right)$ is a function of time $t$ and time separation $\Delta t$. 
	Since the \eqref{equ:ACF_int} includes a definite integral function, the integral part can be calculated by Riemann sum \cite{Jordan2008}, i.e., 
	\begin{equation}
	\label{equ:ACF_sum}
	\tag{18}
	\begin{array}{l}
	C\left( t, t + \Delta t\right) =\\
	\mathrm{E}\left[ {{e^{ - j\frac{{{\omega _{\mathrm{c}}}}}{c}\mathop {\lim }\limits_{N \to \infty } {\mathop{\mathrm Re}\nolimits} \left\{ {\frac{{\Delta t}}{N}\sum\limits_{n = 0}^{N - 1} {a\left( {t + \frac{{n\Delta t}}{N}} \right){e^{j\phi_{0} }}{e^{j{\omega _{\mathrm{v}}}\left( {t + \frac{{n\Delta t}}{N}} \right)}}} } \right\}}}} \right].
	\end{array}
	\end{equation}
	
    \begin{lemma}
		\label{lemmab}
		\rm
		The expression of the $ a\left( {t + \frac{n \Delta t}{N}} \right) $ is 
		\begin{eqnarray}
		\setcounter{equation}{19}
		\label{equ:lemma2}
		&&\hspace{-0.25in}a\left( t +  \frac{n \Delta t}{N} \right)= 
		{e^{  - \mu \frac{n \Delta t}{N}  } }{b_1}\nonumber\\
		&&\hspace{0.6in}+ \sqrt {1 - {e^{  - \mu \frac{ 2\Delta t}{N}}}} \mathop \sum \limits_{k = 2}^{n + 1} {e^{  - \mu \frac{(n + 1 - k) \Delta t}{N}}}{b_k}.
		\end{eqnarray}
	\end{lemma}

	\begin{IEEEproof}
		See Appendix C.
	\end{IEEEproof}

	According to Lemma \ref{lemmab}, the stationary assumption is used to derive the close-from channel temporal ACF expression.
	It leads to the expression of $C\left( t, t + \Delta t\right)$ can be simplified to \eqref{equ:expression}, which is shown at the top of this page, by replacing $a\left( {t + \frac{{n\Delta t}}{N}} \right)$ in \eqref{equ:ACF_sum}. 
	$b_{1}$ and $b_{k}$ in \eqref{equ:expression} are random parameters obey independent and identically distributed (i.i.d.) Gaussian distribution. 
 
	Then, according to characteristic function $\mathrm{ E}\left[ {{e^{jt_cX_c}}} \right] = {e^{jt_c\mu_c  - \frac{{\sigma_{c} ^2{t_c^2}}}{2}}}$, where $t_c$ is the argument of the characteristic function, $X_c$ is the random variable obeys Gaussian distribution, $\mu_c$ is the mean of $X_c$, and $\sigma_c^2$ is the variance of $X_c$, the $C\left( \Delta t \right)$ in \eqref{equ:expression} can be computed \cite{Oberhettinger1973}. 
	In our case, $b_k$ is $X_c$, $\mu_c$ is 0, and $ t_c $ is 1. Therefore, the channel temporal ACF can be simplified as 
	\begin{equation}
	\label{equ:CF}
	\tag{21}
	C\left(t, t+ \Delta t\right)= {e^{-0.5{\sigma_{\mathrm{v}}^2}{\sigma _{X}^2\left( {t, t+\Delta t} \right)}}},
	\end{equation}
	where
	\begin{equation}
	\label{equ:Simga}
	\tag{22}
	\sigma _{X}^2\left( {t, t+\Delta t} \right) = {\left( {\frac{{{\omega _{\mathrm{c}}}}}{c}} \right)^2}\left( \alpha + \beta \right). 
	\end{equation}
	
	Therein, $ \alpha$ and $\beta $ are computed by \eqref{equ:A} and \eqref{equ:B}, which are shown at the top of this page, respectively.
	
	According to definite-integral notation in \cite{Jordan2008}, the \eqref{equ:A} and \eqref{equ:B} can be equaled to \eqref{equ:A_int} and \eqref{equ:B_int}, which are shown at the top of this page, where the variable $n$ in \eqref{equ:A} is replaced by $\mathrm{d}x$ in \eqref{equ:A_int}, and the variable $k$ and $n$ in \eqref{equ:B} are replaced by $\mathrm{d}x_1$ and $\mathrm{d}x_2$ in \eqref{equ:B_int}, respectively. 
	
	The results of \eqref{equ:A_int} and \eqref{equ:B_int} are computed and substituted into \eqref{equ:Simga}, the results are described in \eqref{equ:key}, which is shown at the top of this page.
	It can be seen that $ \sigma_{X}^2\left( \Delta t\right)  $ is a function of $ \phi $, $ \mu$, $\omega_{\mathrm{v}} $,  $\omega_{\mathrm{c}} $, $c $, and $ \Delta t $. The $\omega_{\mathrm{v}} t $ and $ \phi_{0} $ are set to $ \phi $ in \eqref{equ:key}, where $\omega_{\mathrm{v}} t+\phi_{0}=\phi $. 
	In \eqref{equ:End}, which is shown at the top of this page, the $ 2\phi+\omega_{\mathrm{v}} \Delta t $ is set to $ \phi'$, which can simplify the computation. Then, substituting \eqref{equ:End} into \eqref{equ:CF}, the channel temporal ACF closed-form can be computed by using Bessel function \cite{Jordan2008}. The closed-form result is shown in \eqref{equ:bessel}.

%

	\subsection{Proof of Lemma \ref{lemmab} }
	\begin{figure*}[t]
			\vspace{-0.1cm}
		\normalsize
		\begin{eqnarray}
		\setcounter{equation}{31}
		\label{equ:proof_3}
 		&&R\left(\Delta t \right)= {\mathrm{E}}\left[ {{e^{ - \mu \frac{{\left( {{n_1} + {n_2}} \right)\Delta t}}{N}}} + \left( {1 - {e^{ - \mu \frac{{2\Delta t}}{N}}}} \right)\left( {\sum\limits_{k = 2}^{{n_1} + 1} {{e^{ - \mu \frac{{\left( {{n_1} + 1 - k} \right)\Delta t}}{N}}}{b_k}} } \right)\left( {\sum\limits_{k = 2}^{{n_1} + 1} {{e^{ - \mu \frac{{\left( {{n_2} + 1 - k} \right)\Delta t}}{N}}}{b_k}} } \right)} \right] \nonumber\\
		&&\hspace{0.45in}= {\mathrm{E}}\left[ e^ { { - \mu \frac{{\left( {{n_1} + {n_2}} \right)\Delta t}}{N}}  }+ \left( {1 - e^{ { - \mu \frac{{2\Delta t}}{N}} } } \right)\left( {\sum\limits_{k = 2}^{{n_1} + 1} {e^{ - \mu \frac{{\left( {{n_1} + {n_2} + 2 - 2k} \right)\Delta t}}{N}} } }  \right) \right].
		\end{eqnarray} 
		\hrulefill	 
	\end{figure*}

		If the \eqref{equ:lemma2} is correct, according to Assumptions 4, the autocorrelation value of $a\left( t +  \frac{n_1 \Delta t}{N} \right)  $  and $ a\left( t +  \frac{n_2 \Delta t}{N}\right) $ at $ n_1 < n_2 $ should be computed as	
		\begin{equation}
		\label{equ:proof_2}
		\tag{29}
		\begin{array}{l}
		R\left(\Delta t \right)= \mathrm{ E}\left[ a\left( t +  \frac{n_1 \Delta t}{N} \right) a^*\left( t +  \frac{n_2 \Delta t}{N} \right) \right] \\
		\hspace{0.45in}= e^{-\mu \frac{\left( n_2 -n_1 \right) \Delta t}{N}}.
		\end{array}
		\end{equation}
		
		Since the $ b_k $ in \eqref{equ:lemma2} are a series of i.i.d. random variable, so the expectation of $ b_k $ can be computed as
		\begin{equation}
		\tag{30}	
		\label{equ:variable}		
		\mathrm{E}\left[ b_{k_1}\times b_{k_2} \right]=
		\left\{
		\begin{array}{ll}
		1,&k_1=k_2,\\
		0,&k_1\neq k_2.
		\end{array}
		\right.
		\end{equation}
		
	 	According to \eqref{equ:variable}, the $R\left(\Delta t \right)$ can be computed as \eqref{equ:proof_3}, which is shown at the top of this page. Then, the result in \eqref{equ:proof_2} can be computed easily by using geometric sequence property and straightforward mathematics based on \eqref{equ:proof_3}.
\bibliographystyle{IEEEtran}
	\bibliography{References.bib}
\end{document}